\documentclass[aps,prb,superscriptaddress,preprint]{revtex4}
\usepackage{amsmath}
\usepackage{amssymb}
\usepackage{graphicx}

\begin{document}

%

\makeatletter
 \renewcommand\@biblabel[1]{#1.}
\makeatother

\title{Odd and even Kondo effects from emergent localisation\\ in quantum point contacts}

\author{M.~J.~Iqbal}
\affiliation{Zernike Institute for Advanced Materials, University of Groningen, NL-9747AG  Groningen, The Netherlands}
\author{Roi Levy}
\affiliation{Department of Physics, Ben-Gurion University of the Negev, Beer Sheva 84105, Israel}
\author{E.~J.~Koop}
\affiliation{Zernike Institute for Advanced Materials, University of Groningen, NL-9747AG  Groningen, The Netherlands}
\author{J.~B.~Dekker}
\author{J.~P.~de~Jong}
\author{J.~H.~M.~van~der~Velde}
\affiliation{Zernike Institute for Advanced Materials, University of Groningen, NL-9747AG  Groningen, The Netherlands}
\author{D.~Reuter}
\author{A.~D.~Wieck}
\affiliation{Angewandte Festk\"{o}rperphysik, Ruhr-Universit\"{a}t Bochum, D-44780 Bochum, Germany}

\author{R.~Aguado}
\affiliation{Instituto de Ciencia de
Materiales de Madrid (ICMM), Consejo Superior de Investigaciones
Cient\'{\i}ficas (CSIC),
Sor Juana Ines de la Cruz 3, 28049 Madrid, Spain}

\author{Yigal~Meir}
\affiliation{Department of Physics and the Ilse Katz Institute for Nanoscale Science and Technology, Ben-Gurion University of the Negev, Beer Sheva 84105, Israel}

\author{C.~H.~van~der~Wal}
\affiliation{Zernike Institute for Advanced Materials, University of Groningen, NL-9747AG  Groningen, The Netherlands}

\date{Version of \today}

\maketitle


\textbf{A quantum point contact (QPC) is a very basic nano-electronic device: a short and narrow transport
channel between two electron reservoirs. In clean channels electron transport is ballistic and the conductance $G$ is then quantised as a function of channel width  \cite{vanWees1988,Wharam1988} with plateaus at
integer multiples of $2e^2/h$ ($e$ is the electron charge and $h$
Planck's constant). This can be understood in a picture where the
electron states are propagating waves, without need to account for electron-electron interactions.
Quantised conductance could thus be the signature of ultimate control over nanoscale electron transport. However, even studies with the cleanest QPCs generically show significant anomalies on the quantised conductance traces and there is consensus that these result from electron
many-body effects \cite{Thomas1996,Micolich2011}. 
Despite extensive experimental and theoretical studies
\cite{Micolich2011,Cronenwett2002,Meir2002prl,Rejec2006,Koop2007,Komijani2010,Burke2012,Wu2012}
understanding of these anomalies is an open problem. 
We report evidence that the many-body effects have their origin in one or more spontaneously localised states that emerge from Friedel oscillations in the QPC channel. Kondo physics will then also contribute to the formation of the many-body
state with Kondo signatures that reflect the parity of the number of localised
states. Evidence comes from experiments with length-tunable QPCs that show
a periodic modulation of the many-body physics with Kondo signatures of alternating parity. 
Our results are of importance for assessing the role of QPCs in more complex hybrid devices \cite{DelftMajorana,MarcusMajorana} and proposals for spintronic and quantum information applications \cite{Bertoni2000,Blaauboer2005}. In addition, our results show that tunable QPCs offer a rich platform for investigating many-body effects in nanoscale systems, with the ability to probe such physics at the level of a single site.} 

There are two signatures of many-body physics that are generically
observed for a wide variety of QPCs, which includes systems in GaAs \cite{Thomas1996,Komijani2010}, Si \cite{Bagraev2004}, graphene
\cite{Tombros2011}, and both for electron and hole transport
\cite{Komijani2010,Bagraev2004}. First, the quantised conductance traces often show the so-called 0.7~anomaly: an additional small plateau at about $0.7 \cdot (2e^2/h)$. Second, as a function of bias voltage across the channel $G$ typically shows a peak around zero bias (hence named Zero-Bias Anomaly, ZBA), mostly below the first quantised plateau. The experimental
observation \cite{Cronenwett2002} that the ZBA and the 0.7~anomaly
had similarities with the Kondo effect for quantum dots (transport
through a single localised electron state \cite{Goldhaber1998,Cronenwett1998science}),
inspired theoretical work \cite{Meir2002prl,Rejec2006} that
proposed that electron many-body physics could lead to localised electrons in the centre of the QPC. This is a remarkable phenomenon since a QPC is a fully open quantum system. To avoid confusion with localisation by an atomic impurity or disordered potential these many-body states are coined self-consistent or emergent localised states (ELSs).

\begin{figure}[h!]
\centering
\includegraphics[width=8.8cm]{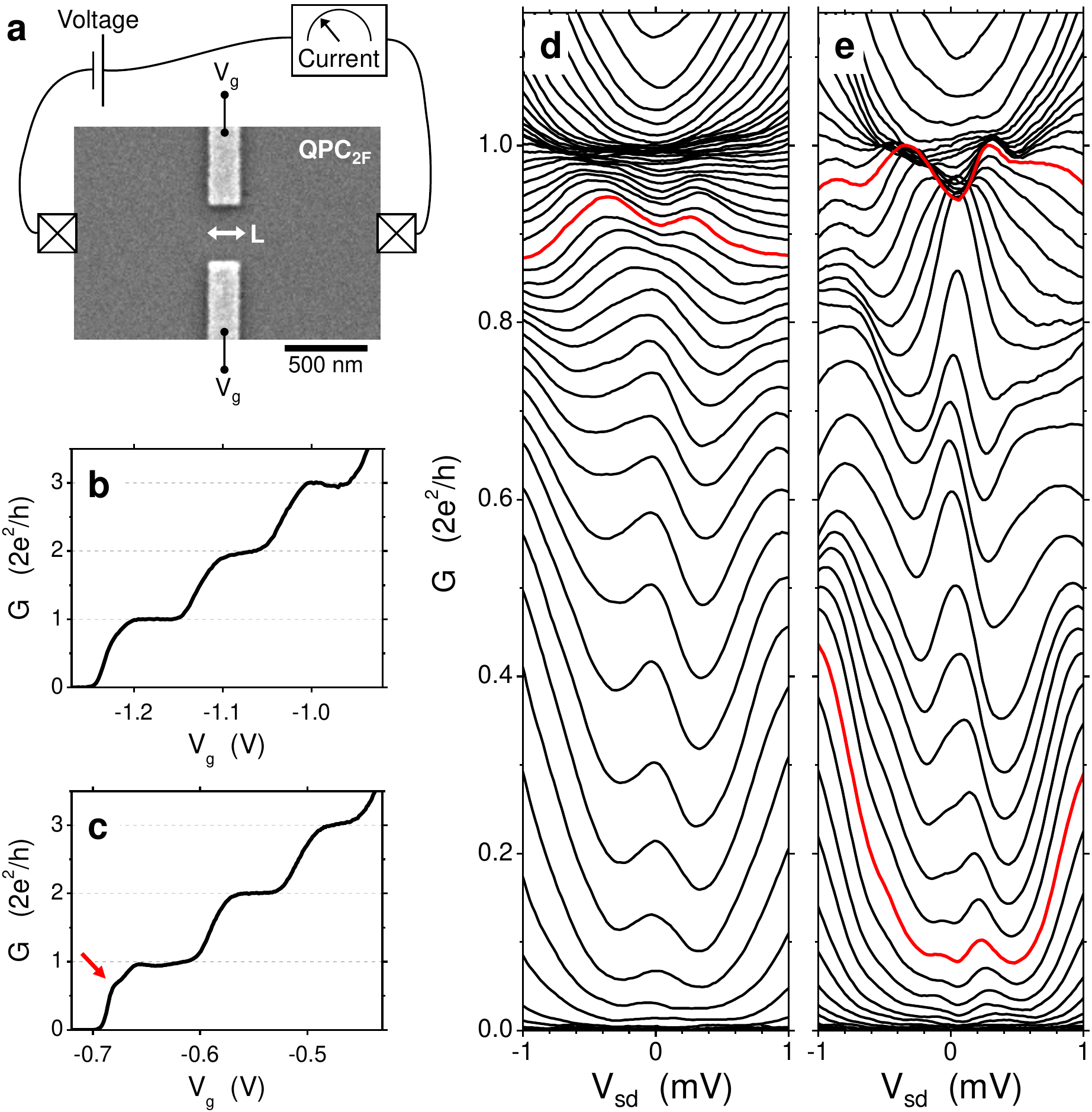}
\caption{\textbf{Conductance of Quantum Point Contacts (QPC).} \textbf{a},
Electron microscope image of a conventional QPC with 2 gate fingers
(QPC$_{{\rm 2F}}$). These gates are on the surface of a wafer with a
two-dimensional electron gas (2DEG) at 110~nm or 114~nm depth.
Applying voltages $V_g$ to these gates induces a narrow transport
channel between source and drain regions of the 2DEG. The
conductance of the QPCs is studied by applying and measuring voltage
and current signals on contacts to the 2DEG. For QPC$_{{\rm 2F}}$
the length of the transport channel is fixed by the lithographic
length $L$. \textbf{b},\textbf{c}, Linear conductance $G$ measured on two
different QPC$_{{\rm 2F}}$ with $L=200~{\rm nm}$. The traces show
clear quantised conductance plateaus at integer multiples of
$2e^2/h$ (the shift in dependence on $V_{g}$ for \textbf{c} as compared to
\textbf{b} is due to a different cool-down procedure, see Methods).
The plateaus and transitions between plateaus show small deviations
from clean quantised conductance behavior, as for example an
additional shoulder at $G\approx 0.7 \cdot(2e^2/h)$ in panel \textbf{c}
(red arrow). \textbf{d}, Nonlinear conductance $G$ as a function of
source-drain voltage $V_{sd}$ at various $V_{g}$ settings, for the
device of panel \textbf{b}. The Zero-Bias Anomaly (ZBA, enhanced
conductance around $V_{sd}=0$) has mostly single-peak character, but
has double-peak character for $G\approx 0.9 \cdot(2e^2/h)$ (for
example the red trace). \textbf{e}, Similar results as panel \textbf{d}
for the device of panel \textbf{c}. In this device ZBAs with double-peak
character appear at $G\approx 0.1 \cdot(2e^2/h)$ and
$G\approx 0.95 \cdot(2e^2/h)$.} \label{Fig:FigQPC2F}
\end{figure}

\begin{figure}[h!]
\centering
\includegraphics[width=\columnwidth]{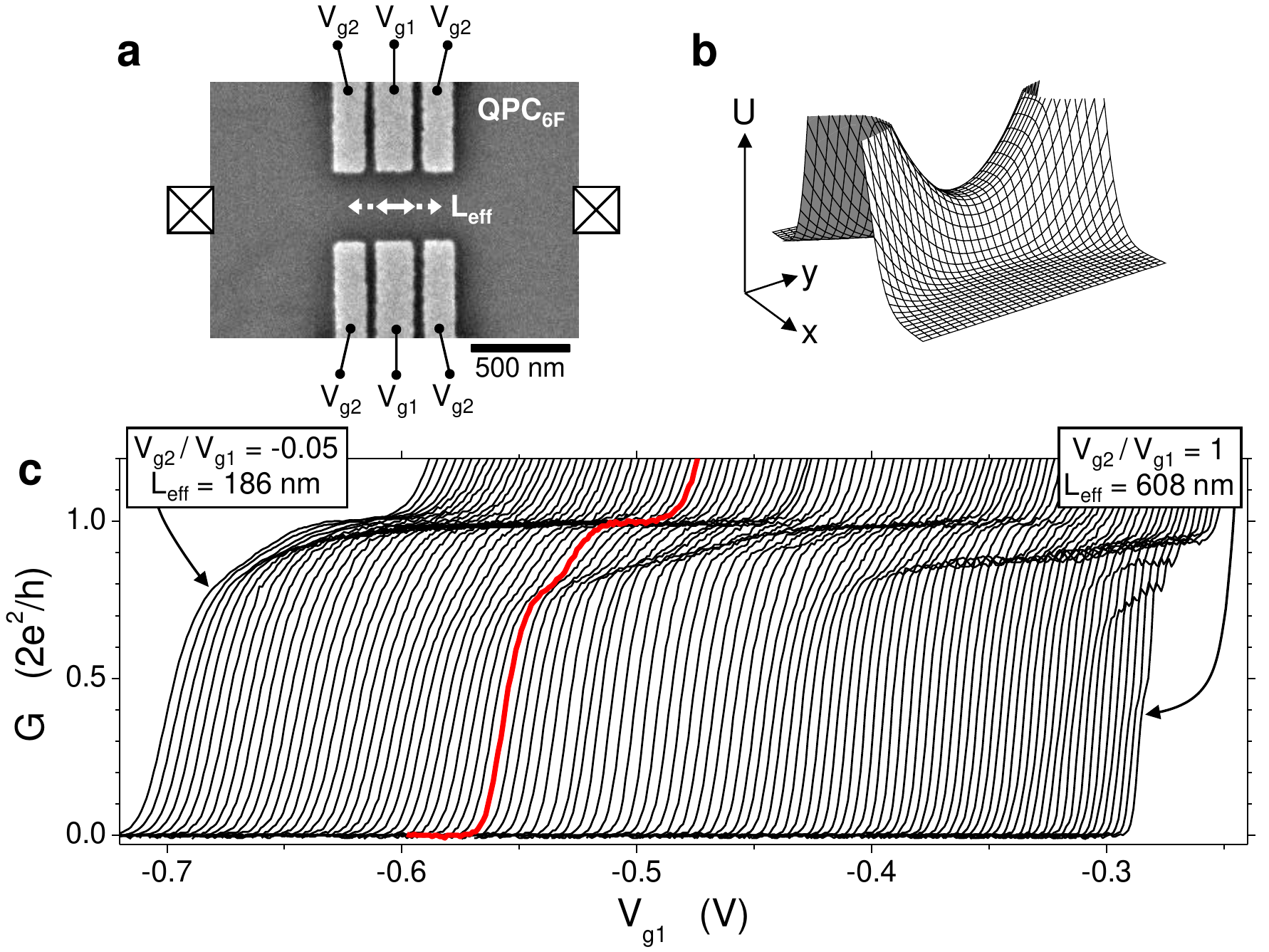}
  \caption{\textbf{Length-tunable quantum point contact (QPC).}
\textbf{a}, Electron microscope image of a QPC with 6 gate fingers (QPC$_{{\rm 6F}}$). It has a tunable effective length $L_{eff}$ that is set by operating at a fixed ratio $V_{g2}/V_{g1}$.
\textbf{b}, Saddle-point potential that illustrates the electron potential energy $U$ (without many-body interactions) in the 2DEG plane in a QPC region.
\textbf{c}, Linear conductance $G$ as a function of $V_{g1}$ (while co-sweeping $V_{g2}$ at fixed $V_{g2}/V_{g1}$) measured on a QPC$_{{\rm 6F}}$ for $L_{eff}$ tuned from 186~nm to 608~nm (traces \textit{not} offset). Besides the quantised conductance plateau at $1 \cdot(2e^2/h)$, most traces show a smaller plateau in the range 0.7 to $0.9 \cdot (2e^2/h)$ due to many-body effects. For this signature 3 periods of modulation can be observed in its dependence on $V_{g2}/V_{g1}$ (\textit{i.e.} $L_{eff}$) .}
\label{Fig:FigLtuneQPC}
\end{figure}

This theoretical work \cite{Meir2002prl,Rejec2006}
developed the picture that the many-body effects in QPC channels are intimately related with the occurrence of a Friedel oscillation  --an oscillation in the electron charge density that occurs when electron waves get reflected in a partially open QPC channel-- which gets enhanced into an ELS with about one electron of charge due to Coulomb repulsion and exchange interactions between
electrons. This reduces the conductance and can explain the 0.7~anomaly \cite{Meir2002prl}. However, transport through such a state can be enhanced by the Kondo
effect at temperatures below a typical Kondo temperature $T_K$. This appears as a ZBA and also moves the 0.7~plateau towards
unity (in units of $2e^2/h$), consistent with experiments.
This theoretical work \cite{Meir2002prl,Rejec2006} also predicted that, depending on parameters, a pair of such ELSs may emerge in the channel, resulting in a double-peak ZBA (as observed in double quantum dots \cite{Jeong2001} due to the two-impurity Kondo effect \cite{Jones1989b,Ivanov1997,Pohjola1997,Aono1998,Georges1999,Aguado2000}). To date no such double-peak ZBAs were reported for QPCs (Supplementary Information Sec.~\ref{Sec:SOMotherReports}). We report here the observation of  such double-peak ZBAs in a large number of conventinal QPCs (with 2 gate fingers as in Fig.~\ref{Fig:FigQPC2F}a, denoted as QPC$_{{\rm 2F}}$). We also introduce a new type of QPC which has a tunable channel length (with 6 gate fingers as in Fig.~\ref{Fig:FigLtuneQPC}a, denoted as QPC$_{{\rm 6F}}$). In these devices the 0.7~anomaly and ZBA show a periodic modulation as a function of QPC length, which we attribute to an increasing number of ELSs. Thus, as the number of ELSs increases with QPC length, its parity alternates, giving rise to modulation between odd and even-impurity Kondo effects, and, as a result, between single- and double-peak ZBA. In addition, the 0.7~anomaly shows a periodic modulation because the enhancement of the 0.7~feature towards unity conductance depends on both the parity and the parameters of the Kondo system such as $T_K$, and these are both modulated as a function of QPC length.


The QPCs were realised by locally depleting the two-dimensional electron gas (2DEG) below the surface of a GaAs/AlGaAs heterostructure (for
details see below, Methods). In the standard approach, applying negative voltage to two metallic gate fingers on the surface of such material (Fig.~\ref{Fig:FigQPC2F}a) induces an electrostatic potential barrier between a source and drain reservoir in the 2DEG, with a small tunable opening in the form of a saddle-point potential (Fig.~\ref{Fig:FigLtuneQPC}b). Such devices with two gate fingers (QPC$_{{\rm 2F}}$) have a fixed channel length $L$. The novel  devices with six gate fingers (QPC$_{{\rm 6F}}$, see Fig.~\ref{Fig:FigLtuneQPC}a) have a channel with tunable
length $L_{eff}$. We focus on the differential conductance $G$,
which is obtained by applying and  measuring voltage and current
signals as in the simplified scheme in Fig.~\ref{Fig:FigQPC2F}a
(for details see below, Methods). Results for $G$ at zero
bias voltage $V_{sd}$ will be called linear conductance, while
results for $G$ as a function of $V_{sd}$ (bias spectroscopy) will
be called nonlinear conductance. Unless stated otherwise, we present
data taken at a temperature $T$ of 80~mK and zero magnetic field.



The signatures of a pair of ELSs are more likely to be observed on shorter QPCs \cite{Rejec2006}. We thus focussed on QPC$_{{\rm 2F}}$ of lithographic length $L = 200~{\rm nm}$ (and width $W = 350~{\rm nm}$), shorter than most QPCs in the literature.
We searched for double-peak ZBAs in a set of 80 QPC$_{{\rm 2F}}$ (realised in two different wafer materials, different fabrication runs, different cool-downs, and with or without gate-biased cool-down, see Methods) and found them in about half the studied devices. The ubiquity of the phenomenon, and the fact that such double-peak ZBAs were persistently observed in the same devices over different cool-downs, implies that it is a generic effect and not due to a fortuitous impurity nearby.
Figure~\ref{Fig:FigQPC2F} presents data from two QPC$_{{\rm 2F}}$ to illustrate that the signatures of many-body physics show qualitatively similar features, though with significant device-to-device variation
(while there is no strong variation in the manifestation of non-interacting electron physics such as the quantised conductance \cite{Koop2007,Burke2012}).
Figures~\ref{Fig:FigQPC2F}b,c present measurements of the linear conductance. In addition to the quantised conductance plateaus at integer multiples of $2e^2/h$, the trace in panel \textbf{c} shows an additional shoulder at $G\approx 0.7
\cdot (2e^2/h)$ (the 0.7~anomaly, also observed for the device of panel \textbf{b} at higher
temperatures). Results for the nonlinear
conductance for these same devices are presented in
Fig.~\ref{Fig:FigQPC2F}d,e.
Most traces between 0 and $1
\cdot(2e^2/h)$ show a single-peak ZBA (the enhancement of
conductance within $\pm$0.5~mV around $V_{sd}=0~{\rm mV}$). However, the red traces mark examples where the ZBA appears with double-peak
character (the asymmetric character of these nonlinear conductance
traces will be discussed below). For double-peak ZBAs just below $1
\cdot(2e^2/h)$ we typically observe a peak splitting that increases with conductance.

The theoretical work \cite{Rejec2006} predicted that the Friedel oscillation from screening the QPC potential creates two electron puddles on the two sides of the QPC. For most QPC geometries it showed that lowering the potential by gate voltage gives a single ELS in the center of the QPC since the Friedel oscillations from both sides connect. However, for short QPCs (for $L$ similar to the Fermi wavelength) the two ELSs remain intact as the potential is lowered. This gives the physics of a two-impurity Kondo system and, as the coupling between them gets stronger with increasing conductance the ZBA splits \cite{Ivanov1997,Pohjola1997,Aono1998,Georges1999,Aguado2000}. At lower conductance values, the ZBA may appear as single- or double-peak depending on the ratio between the Kondo temperature of the two ELSs and the strength of interaction between them. While this is fully consistent with our observations (Fig.~\ref{Fig:FigQPC2F}d,e), the coupling between these ELSs and the resulting splitting of the ZBA depends very sensitively on parameters and the always present remote imperfections (Supplementary Information Sec.~\ref{Sec:SOMremoteImp}), and this leads to significant variability among devices. Thus, a more stringent test of this picture is possible with QPCs whose parameters can be modified continuously.

\begin{figure}[h!]
\centering
\includegraphics[width=\columnwidth]{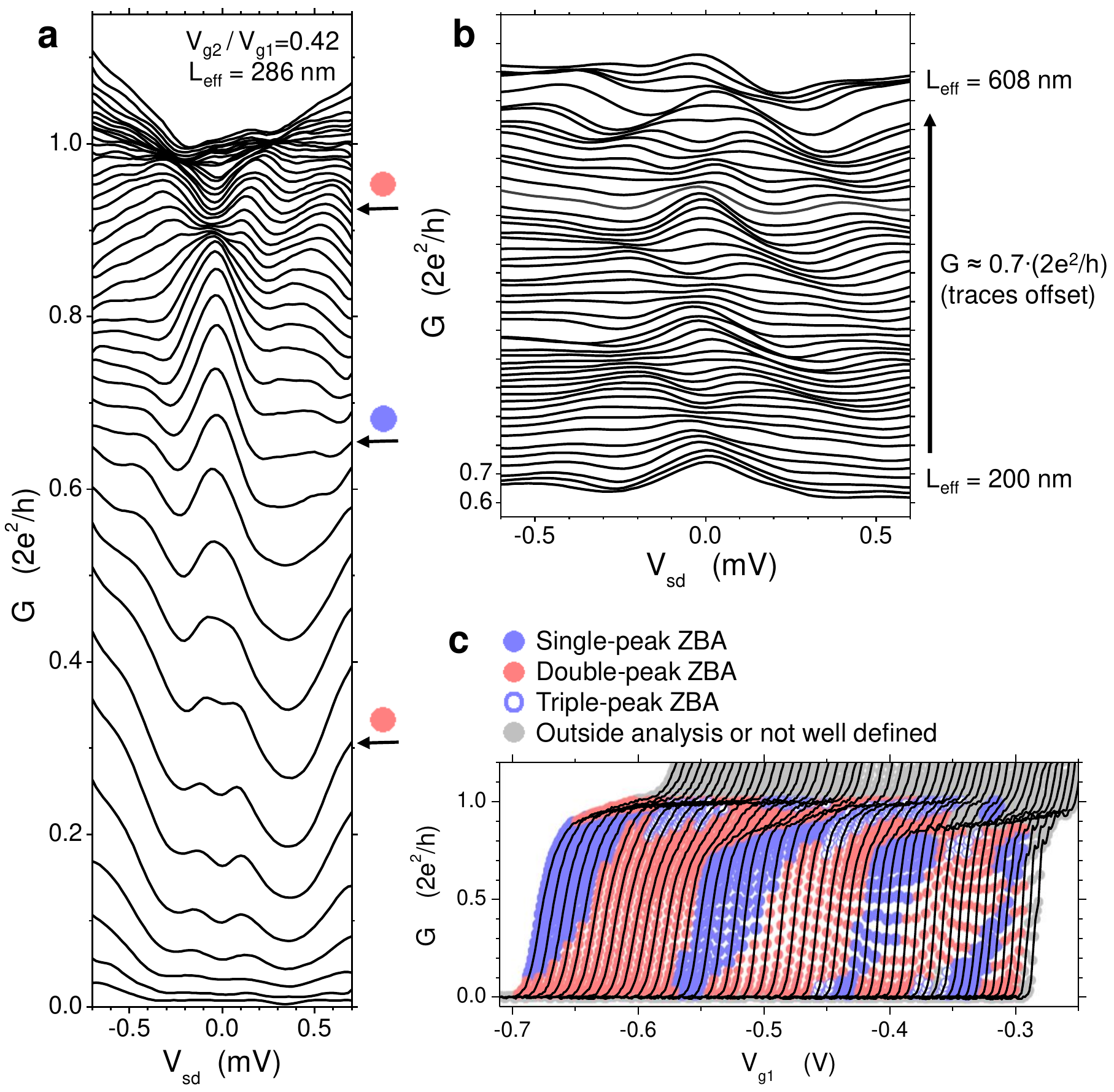}\\
\caption{\textbf{Zero-Bias Anomalies (ZBAs) in the nonlinear conductance of
a QPC$_{{\rm {\bf 6F}}}$.} \textbf{a}, Nonlinear conductance $G$ as a
function of source-drain voltage $V_{sd}$ at various $V_{g1}$
settings, for operation at $L_{eff}=286~{\rm nm}$. The ZBA appears
alternatingly with single- or double-peak character. \textbf{b},
Evolution of the ZBA in the nonlinear conductance at fixed
conductance level of $\sim 0.7 \cdot (2e^2/h)$ as a function of
$L_{eff}$ (traces offset). The ZBA has alternatingly single- or
double-peak character. \textbf{c}, The character of the ZBA
(single-, double-, or triple-peak, as labeled) mapped out on the
linear conductance data of Fig.~\ref{Fig:FigLtuneQPC}c.}
\label{Fig:FigZBAcharact}
\end{figure}

We addressed this by measuring a set of 8 QPC$_{{\rm 6F}}$
devices, for which we could gradually change the effective
length $L_{eff}$ of the saddle-point potential. These devices showed qualitatively identical behavior with oscillatory signature of the 0.7~anomaly and regular modulation between single- and double-peak ZBAs with increasing
$L_{eff}$ (Fig.~\ref{Fig:FigZBAcharact}b,c). These reproducible
observations on 8 QPC$_{{\rm 6F}}$ (and for different cool-downs of one QPC$_{{\rm 6F}}$, Supplementary Information Sec.~\ref{Sec:SOMlinGotherQPC6F})
are consistent with the emergence of an increasing number of ELSs due to many-body physics that generically occurs, also in ultra-clean QPC channels.


Figure~\ref{Fig:FigLtuneQPC}a depicts the QPC$_{{\rm
6F}}$ devices, for which the channel length $L_{eff}$ could be tuned
continuously. These were operated with
the central gate voltage $V_{g1}$ more negative than the side gate voltage $V_{g2}$ to
avoid quantum dot formation.  We analyzed that in this regime the gates induce a smooth saddle-point potential, despite the narrow gaps between the gate fingers. The effective length $L_{eff}$ is set by $V_{g2}/V_{g1}$ (short for
$V_{g2}/V_{g1}$ near 0, long for $V_{g2}/V_{g1}$ near 1, see Supplementary Information Sec.~\ref{Sec:SOMcalLeff} for details). Our devices could thus be controlled to have $L_{eff}$ from about 186~nm to 608~nm. Making
$V_{g1}$ less negative, at fixed ratio $V_{g2}/V_{g1}$, opens the QPC while keeping the length unchanged.

All QPC$_{{\rm 6F}}$ showed clear quantised
conductance plateaus. Figure~\ref{Fig:FigLtuneQPC}c shows for one device how the ``0.7~anomaly'' appears as
a smaller plateau in the range 0.7 to $0.9 \cdot
(2e^2/h)$, which shows a dependence on $L_{eff}$ with 3
periods of modulation for the range
$L_{eff}=186~{\rm nm}$ to 608~nm. The nonlinear conductance measurements
from this device in Fig.~\ref{Fig:FigZBAcharact}a show how the ZBAs appear for
$L_{eff} = 286~{\rm nm}$. At fixed length, the ZBAs alternate
between single-peak and double-peak character when opening the QPC, again with increased splitting for the double-peak ZBA as the conductance approaches $1
\cdot(2e^2/h)$. The overall appearance of the ZBAs is very similar that of fixed-length QPC$_{{\bf {\rm 2F}}}$ (Fig.~\ref{Fig:FigQPC2F}d,e).
Figure~\ref{Fig:FigZBAcharact}b shows that there is also a
modulation between single- and double-peak character when increasing
$L_{eff}$ at a fixed conductance level (as evaluated
immediately next to the ZBA). Figure~\ref{Fig:FigZBAcharact}c plots
again the data of Fig.~\ref{Fig:FigLtuneQPC}c, with colored symbols
on the traces that mark whether the ZBA at that point has single- or
double-peak character (in some cases we find ZBAs that are best
described as triple-peak). The modulation between single-peak and
double-peak ZBA as a function of $L_{eff}$ also shows about 3
periods, and is clearly correlated with the modulation of the 0.7~anomaly.


To critically check the relevance of two-impurity Kondo physics for our observations we measured the temperature and magnetic-field dependence of double-peak ZBAs and compare this with theory for this system. We obtained the theoretical results from calculating the current through a two-impurity Anderson model.
For this model one expects \cite{Ivanov1997,Pohjola1997,Aono1998,Georges1999,Aguado2000} a current that gets enhanced by the Kondo effect when lowering the temperature from above to below the Kondo temperature. Depending on the strength of the effective coupling between the impurities relative to the Kondo temperature the associated ZBA has either single- or double-peak character.

\begin{figure}[h!]
\centering
\includegraphics[width=12cm]{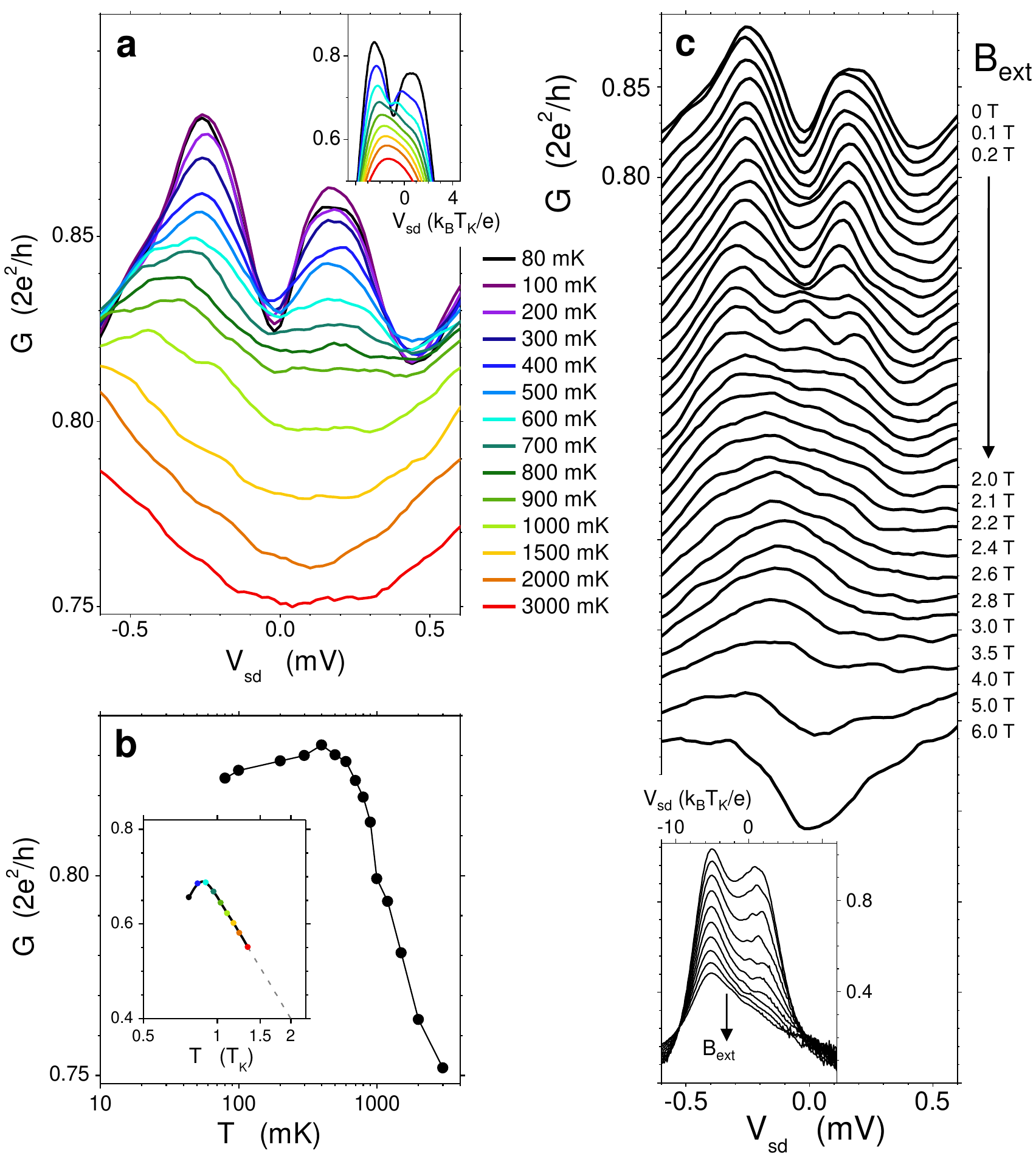}
\caption{\textbf{Comparison between experiments and theory of the Anderson model for a two-impurity Kondo system.}
\textbf{a}, Evolution of the nonlinear conductance (with double-peak ZBA) as a function of temperature for a QPC$_{{\rm 6F}}$ operated with fixed $V_{g1}=-0.528~{\rm V}$ and fixed $V_{g2}/V_{g1}=0.3$ (traces not offset). The inset presents calculated nonlinear-conductance traces from a two-impurity Kondo model (see main text).    
\textbf{b}, The conductance level of panel \textbf{a} near zero bias (between the two ZBA peaks, evaluated at $V_{sd}=-0.02~{\rm mV}$) as a function of temperature (logarithmic axis).
The inset presents the same analysis applied to the theoretical results in the inset of panel \textbf{a}.
\textbf{c}, Evolution of the nonlinear conductance (same device and operation point as panel \textbf{a}) as a function of applied in-plane magnetic field $B_{ext}$ (traces offset $-0.01 \cdot (2e^2/h)$). The inset presents again calculated nonlinear-conductance traces from a two-impurity Kondo model.}
\label{Fig:FigTBdep}
\end{figure}

Figure~\ref{Fig:FigTBdep}a depicts how the nonlinear conductance develops from a background conductance of $0.75 \cdot (2e^2/h)$ at 3000~mK into a double-peak ZBA with peak values up to about $0.90 \cdot (2e^2/h)$ as the temperature is decreased (device of Fig.~\ref{Fig:FigZBAcharact}, Supplementary Fig.~\ref{Fig:FigTdepLIN} shows temperature dependence of the linear conductance).
The conductance between the peaks (the linear
conductance) has a nonmonotonic temperature
dependence (Fig.~\ref{Fig:FigTBdep}b) that is characteristic for two-impurity Anderson physics \cite{Aguado2003}.
Insets in Fig.~\ref{Fig:FigTBdep}a,b depict for comparison results of the theoretical calculations and show good qualitative agreement.
(Note that this description only calculates the Kondo contribution to the current which yields zero current at high temperatures due to Coulomb blockade, while the ELSs in QPCs are not expected to show strong Coulomb blockade at high temperatures but a finite background conductance, as observed.)
The theoretical traces are calculated for two impurity sites with unequal coupling strength $\Gamma$ to a neighboring electron reservoir, and accordingly an unequal Kondo temperature $T_K$ (the plots are presented on the energy scale given by the highest Kondo temperature of the two sites). The asymmetry between the two $\Gamma$ parameters gives asymmetric double-peak ZBAs, with the
minimum conductance between the peaks not exactly at
$V_{sd}=0$, very similar to the  experimentally observed double-peak ZBAs. We should expect such asymmetries between the two $\Gamma$ parameters since the $\Gamma$ values depend exponentially on the coupling between the ELSs and the electron reservoirs (Supplementary Information Sec.~\ref{Sec:SOMremoteImp}).


Figure~\ref{Fig:FigTBdep}c presents magnetic-field dependence of this double-peak ZBA. Theoretically, one expects different evolutions with magnetic field depending on the relative magnitude of parameters, such as the Kondo temperature of each impurity, the coupling between the impurities, the asymmetry between the impurities and temperature. The possibilities include: each peak splits into two sub-peaks, the peaks get closer and merge, the peaks move closer and cross each other, etc.
We see such diverse dependence on magnetic field for the different operation points of a device, and for different devices. Figure~\ref{Fig:FigTBdep}c shows one example with a comparison to the theoretical expectations (other examples are in Supplementary Information Sec.~\ref{Sec:SOMdataBdepNONLIN}). We see again good qualitative agreement between the theory of the two-impurity model and the experimental observations.

From the width of the peaks in Fig.~\ref{Fig:FigTBdep}a one can estimate the Kondo temperature $T_K$, while the splitting between the peaks is then equal to twice the effective coupling between the two ELSs. This coupling is here $\sim$0.2~meV (typical value for the larger data set behind Fig.~\ref{Fig:FigZBAcharact}c). This should be consistent with the temperature where the double-peak character of the ZBA is no longer resolved (and the temperature where the nonmonotonic temperature dependence of conductance has a maximum \cite{Aguado2003}).
In Fig.~\ref{Fig:FigTBdep}a this occurs for $\sim$800~mK (\textit{i.e.}
$\sim$0.1~meV). This is in reasonable agreement with the distance
between the peaks. The fact that it is on the low side is probably because $T_K$ is here
at the same energy scale as the splitting ($\sim$0.15~meV when estimated as
half the peak widths in Fig.~\ref{Fig:FigTBdep}a). We also analyzed ZBAs with pronounced single-peak character in the same manner as Cronenwett \textit{et al.}~\cite{Cronenwett2002} and found agreement with the single-impurity Kondo model to the same extent.


The increase of $L_{eff}$ that induces one period of modulation for the 0.7~anomaly and the split-peak character of the ZBA in Fig.~\ref{Fig:FigZBAcharact} ($\sim$100~nm to $\sim$150~nm) matches with the Fermi wavelength in
the QPC channel (Supplementary Information Sec.~\ref{Sec:SOMfriedel}). This supports the hypothesis that the periodicity is linked to Friedel oscillations in the channel which gradually develop additional periods as it gets longer.
(That such Friedel oscillations occur upon scattering in a 2DEG has been observed directly in a different setting  \cite{Kanisawa2001prl}.)
The increasing number of ELSs with channel length leads to alternation between odd and even-impurity Kondo effects, and to modulation between single-peak and double-peak ZBAs, respectively (the three-impurity case was studied in ref.~\onlinecite{Vernek2009} and can show ZBAs with triple- or predominantly single-peak character).

To support this picture we carried out numerical Spin-Density-Functional-Theory (SDFT) calculations, generalizing earlier work \cite{Rejec2006} (Supplementary Information Sec.~\ref{Sec:SOMonSDFT}). The length of the channel was determined by a gate of variable length.
We studied whether the SDFT yields localised states with about one electron of charge as the state with lowest energy.
The results (an example is in Supplementary Fig.~\ref{Fig:FigSDFT}) show an increasing number of ELSs as the channel gets longer, as well as for opening the QPC. These calculations support the picture we have presented here: The number of ELSs increases by one each time when the QPC length increases over a range that allows for one additional period of the Friedel oscillation in the QPC channel.
Interestingly, the SDFT results suggest that, depending on the overlap of the ELSs, higher spin states ($S$=1 or $S$=3/2) may develop in the QPC, making it possible to study Kondo effects for higher spin, and transport through such exotic states.

The emergence of a ZBA for a hybrid device with a semiconductor channel was recently reported as a signature of Majorana fermions \cite{DelftMajorana}. The fact that similar ZBAs occur with rich behavior in plain semiconductor QPCs suggest that one should be cautious when ruling out alternative explanations for these Majorana signatures \cite{MarcusMajorana}. Evidently, basic understanding of the physics in QPCs is a crucial step in understanding more complex hybrid structures.
Tunable QPCs offer an excellent new test ground for studies to this end, while they are also suited for detailed studies of Friedel
oscillations \cite{Simion2005} and strongly correlated electrons in low dimensions, at the level of a single
site.

%
%
%
%


%

\vspace{1cm}

%
%

\vspace{0.4cm}


\noindent\textbf{METHODS}

\vspace{2mm}

\noindent \textbf{Materials and device fabrication}

\noindent QPC devices were
fabricated with two different GaAs/AlGaAs heterostructures
containing a two-dimensional electron gas (2DEG) in a heterojunction quantum well. They had similar
parameters for the molecular-beam-epitaxy growth and properties of the 2DEG. We obtained very
similar results with both materials. Most of the results presented
in this report come from Material 2, only the data in
Fig.~\ref{Fig:FigQPC2F}b,d comes from Material 1.

Material 1 was a ${\rm GaAs}/{\rm Al}_{0.32}{\rm Ga}_{0.68}{\rm As}$
heterostructure with a 2DEG at 114~nm below the surface from
modulation doping with Si at about $1\cdot10^{24}~{\rm m^{-3}}$. At
4.2 K, the mobility of the 2DEG was $\mu = 159 \; {\rm m^{2}/Vs }$,
and the electron density $n_{s} = 1.5 \cdot 10^{15} \; {\rm m^{-2}}$
after cooling down in the dark. The layer with modulation doping
started at 37~nm distance from the 2DEG position towards the wafer
surface (this material was uniquely used in the related results
presented in ref.~\onlinecite{Koop2007} of the main text).

Material 2 was a ${\rm GaAs}/{\rm Al}_{0.35}{\rm Ga}_{0.65}{\rm As}$
heterostructure with a 2DEG at 110~nm below the surface from
modulation doping with Si at about $1\cdot10^{24}~{\rm m^{-3}}$. At
4.2 K, the mobility of the 2DEG was $\mu = 118 \; {\rm m^{2}/Vs }$,
and the electron density $n_{s} = 1.60 \cdot 10^{15} \; {\rm
m^{-2}}$. Here the layer with modulation doping started at 45~nm
distance from the 2DEG position.

QPCs were realised by locally depleting the 2DEG below the surface of the GaAs/AlGaAs heterostructures.
In the standard approach, applying negative voltage $V_g$ to two metallic gate fingers on the surface of such material (Fig.~\ref{Fig:FigQPC2F}a) induces an electrostatic potential barrier between a source and drain reservoir in the 2DEG, with a small tunable opening in the form of a saddle-point potential (Fig.~\ref{Fig:FigLtuneQPC}b). Such devices with two gate fingers are denoted as QPC$_{{\rm 2F}}$ and these have a fixed channel length $L$. We also studied novel devices with six gate fingers (Fig.~\ref{Fig:FigLtuneQPC}a),
denoted as QPC$_{{\rm 6F}}$, which have a channel with tunable
length $L_{eff}$.
Tuning of $V_g$ on the gates allows for
controlling the effective QPC shape (for details see Supplementary Information Sec.~\ref{Sec:SOMcalLeff}).

The depletion gates were
defined with standard electron-beam lithography and lift-off
techniques and by depositing 15~nm of Au with a Ti sticking
layer. The reservoirs were connected to macroscopic leads via ohmic
contacts, which were realized by annealing a thin Au/Ge/Ni layer
that was deposited on the surface. Part of our data (including all
the results presented in the main text, except for the data in
Fig.~\ref{Fig:FigQPC2F}b,d) was obtained after cooling down with
about +0.3~V on the gates for suppressing $1/f$ and telegraph noise
in the conductance signals due to charge instabilities in the doping
layer (Supplementary Information Sec.~\ref{Sec:SOMremoteImp}). We obtained (besides the change in
noise properties) similar results for the cases with and without
biased cool-down.

\vspace{1cm}

\noindent \textbf{Measurement techniques and setup}

\noindent The measurements focus on the differential conductance $G$,
which is obtained by applying and  measuring voltage and current
signals as in the simplified scheme in Fig.~\ref{Fig:FigQPC2F}a. Results for $G$ at zero
bias voltage $V_{sd}$ are called linear conductance, while
results for $G$ as a function of $V_{sd}$ (bias spectroscopy) are called nonlinear conductance. Unless stated otherwise, the presented
data was taken at a temperature $T$ of 80~mK and zero magnetic field.

The presented results of linear and nonlinear conductance
measurements all concern the differential conductance $G=dI/dV_{sd}$
(where $I$ is the measured current). For linear conductance
measurement we used standard lock-in techniques (typically at 387
Hz), with an ac voltage bias $V_{sd}=V_{ac}=10~{\rm \mu V}$. For the
nonlinear conductance measurements we superimposed an ac and a dc
voltage bias, $V_{sd}=V_{dc}+V_{ac}$. We used an effective
4-terminal measurement where we locally measured the source-drain
voltage drop $V_{sd}$ across the QPC, such that we can present
results without significant contributions from series resistance.
Only one of the source-drain contacts was connected to the grounded
shielding of our setup, and all gate voltages were applied with
respect to this ground.

Measurements were performed in a dilution refrigerator with the
sample at temperatures from $\sim 5 \; {\rm mK}$ to 4.2~K. For all
our data the temperature dependence saturated when cooling below
$\sim 80$~mK. This is consistent with independent measurements of
the lowest effective electron temperature that could be achieved
with this setup. The electron temperature of 80~mK allows for probing peak structures in nonlinear conductance traces as narrow as $4 k_B T/e = 0.03$~mV
($k_B$ is Boltzmann's constant).

The in-plane magnetic field was applied both parallel and
perpendicular to the current direction and we measured devices both
with the current along the $[110]$ and $[-110]$ directions of the
crystal, but the results did not depend significantly on these
orientations. Alignment of the sample with the magnetic field was
within 1$^\circ$, as determined from Hall voltage measurements on
the 2DEG.

\vspace{1cm}

\noindent\textbf{Kondo transport calculations}

\noindent We obtained the theoretical results from calculating the current through a two-impurity Anderson model within the slave-boson noncrossing approximation \cite{Aguado2000} (detailed in Supplementary Information Sec.~\ref{Sec:SOM2KondoTheory}).

\vspace{1cm}

\noindent\textbf{SDFT calculations}

\noindent We obtained the SDFT results by extending the work of ref.~\onlinecite{Rejec2006}  (detailed in Supplementary Information Sec.~\ref{Sec:SOMonSDFT}).

\newpage


\newpage


\newpage


\newpage


\newpage

\newpage

\noindent \textbf{REFERENCES FOR THE MAIN TEXT}

\vspace{1cm}

\noindent \textbf{Supplementary Information} is included as an appendix to this document.

\vspace{1cm}

\noindent \textbf{Acknowledgements} We thank B.~J.~van Wees, A.~Aqeel, S.~Ludwig, J.~von Delft and Y.~Komijani for discussions and B.~Wolfs, J. Holstein and M. de Roosz for technical assistance.
We acknowledge financial support from the German programs DFG-SPP
1285, Research school Ruhr-Universit\"{a}t Bochum
and BMBF QuaHL-Rep 16BQ1035, and grants FIS2009-08744 and FIS2012-33521
from the Spanish
Ministry of Economy and Innovation. M.J.I. acknowledges a scholarship from the Higher Education Commission of Pakistan. Y.M. and R.L. acknowledge support from the ISF.

\vspace{1cm}

\noindent \textbf{Author Contributions} M.J.I. was the lead researcher for experiments with C.H.W. as supervisor and experimental contributions from E.J.K., J.B.D., J.P.J. and J.H.M.V. and design contributions from Y.M. The devices were fabricated from wafer material that was grown by D.R and A.D.W. The calculations of electron transport in Kondo systems were carried out by R.A. The SDFT contribution was worked out by R.L. with Y.M. as supervisor.
M.J.I, C.H.W. and Y.M. wrote the paper.

%
%



%
%

\setcounter{figure}{0}

\renewcommand{\thesection}{\arabic{section}}
\renewcommand{\thefigure}{S\arabic{figure}}

\makeatletter
 \renewcommand\@biblabel[1]{#1.}
\makeatother


\newcommand{\be}[1]{\begin{eqnarray}  {\label{#1}}}
\newcommand{\ee}{\end{eqnarray}}

\begin{center}
\textbf{{\LARGE Supplementary Information}}

for

\textbf{{\Large Odd and even Kondo effects from emergent localisation\\ in quantum point contacts}}

by

M. J. Iqbal, Roi Levy, E. J. Koop, J. B. Dekker, J. P. de Jong, J. H. M. van der Velde,\\
D. Reuter, A. D. Wieck, R. Aguado, Yigal Meir, and C. H. van der Wal

\end{center}

\vspace{2cm}


\noindent \textbf{TABLE OF CONTENTS}

\begin{figure}[h!]
\includegraphics[width=\columnwidth]{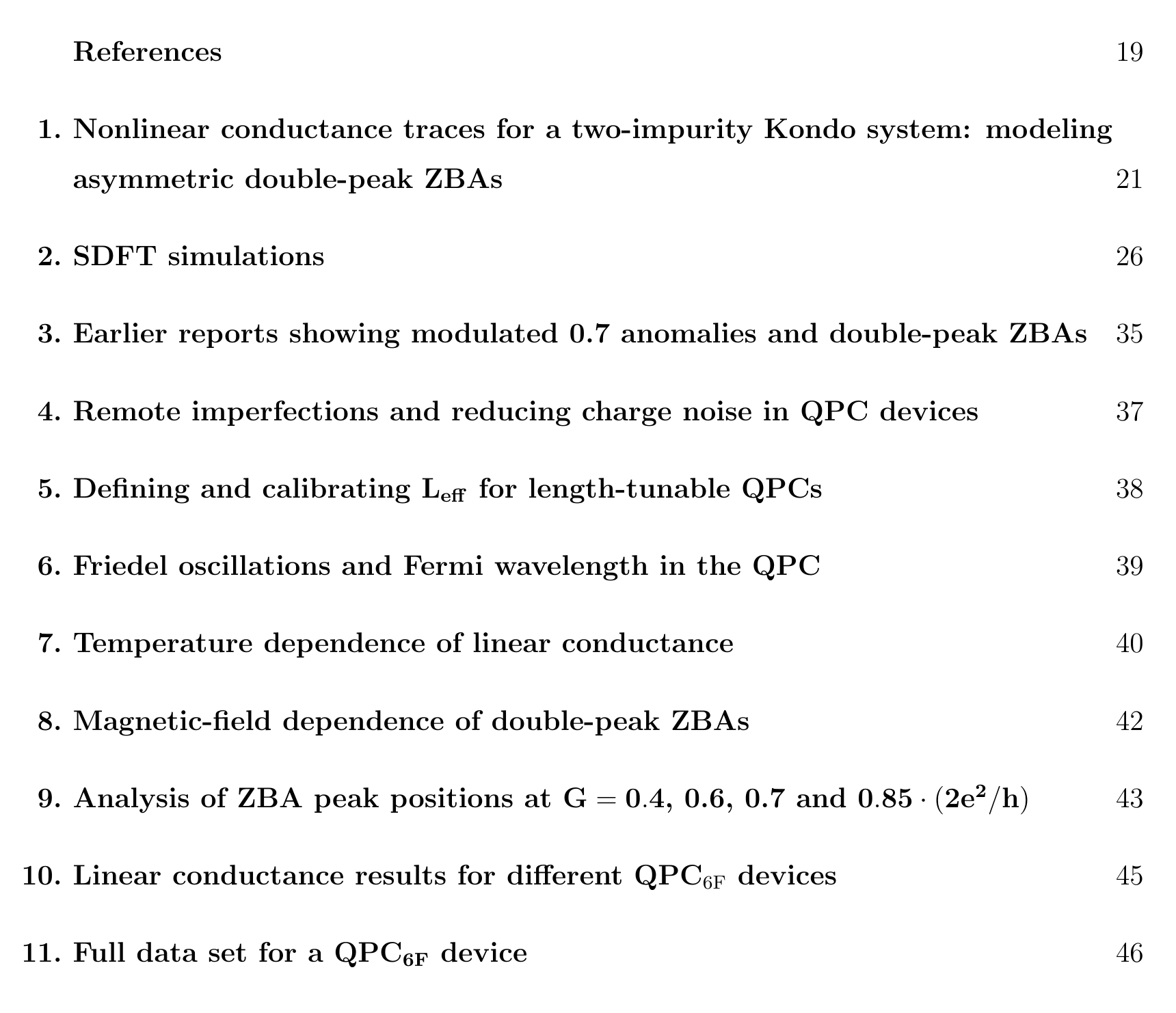}
\end{figure}


\newpage

\noindent \textbf{REFERENCES FOR THE SUPPLEMENTARY INFORMATION}

\vspace{-16mm}


\newpage

\section{Nonlinear conductance traces for a two-impurity Kondo system: modeling asymmetric double-peak ZBA${\rm {\bf s}}$}\label{Sec:SOM2KondoTheory}

This section summarizes the approach that we used for calculating the nonlinear differential-conductance traces of Fig.~\ref{Fig:FigTBdep} in the main text.
The experimental results show nonlinear conductance traces with highly asymmetric double peak ZBAs: the peak height and width of the two peaks can differ significantly, and the minimum between the two peaks can appear at positions that differ significantly from $V_{sd}=0$. The purpose of the theoretical modeling that is presented in this Section is to investigate whether such asymmetric double-peak ZBAs are consistent with the physics of the two-impurity Kondo system.
The theoretical traces of Fig.~\ref{Fig:FigTBdep} are calculated for such a two-impurity Kondo system, and show behavior that is for a large part consistent with the behavior of the experimentally observed asymmetric double-peak ZBAs.
In addition, several devices showed rather symmetric single-peak ZBAs that had a position that differed significantly from $V_{sd}=0$. This behavior also comes forward from the two-impurity Kondo modeling for parameters where the two impurity spins have unequal coupling to their reservoir and for a spin-spin coupling that is weaker than temperature (traces not shown here).

The main challenge for this theoretical modeling is to extend the two-impurity Anderson Hamiltonian to calculations of the differential conductance for transport through two series-coupled impurities at nonzero bias voltage and nonzero temperatures. The approach that we use here directly builds on refs.~\onlinecite{SI-Aguado2000,SI-ram03}. The model parameters that were used for calculating the traces in Fig.~\ref{Fig:FigTBdep} of the main text are summarized at the end of this Section.

\vspace{0.5cm}

\noindent \textbf{Theoretical model}

\noindent
The system is modeled as a
two-impurity Anderson Hamiltonian with an extra term accounting for inter-impurity coupling.
Each impurity is connected to a different Fermi sea with chemical potential
$\mu_L=\frac{eV_{sd}}{2}$ and $\mu_R=-\frac{eV_{sd}}{2}$, respectively. The full Hamiltonian reads
\begin{widetext}
\begin{eqnarray}
H&=&\sum _{ k_{\alpha\in \{ L,R \}}, \sigma} \epsilon _{k_\alpha} c^\dagger_{k_\alpha,\sigma}
c_{k_\alpha,\sigma}
+ \sum _{\alpha\in \{ L,R \},\sigma} \epsilon _{\alpha\sigma} d^\dagger_{\alpha\sigma}
d_{\alpha\sigma}
+V_0\sum _{k_{\alpha\in \{ L,R \}},\sigma }(c^\dagger_{k_\alpha, \sigma} d_{\alpha\sigma}
+ d^\dagger_{\alpha\sigma} c_{k_\alpha, \sigma})\nonumber\\
&+&V_I\sum _\sigma(d^\dagger_{L\sigma}d_{R\sigma}+d^\dagger_{R\sigma}d_{L\sigma})
+U_L n_{L,\uparrow} n_{L,\downarrow}
+U_R n_{R,\uparrow} n_{R,\downarrow}.
\end{eqnarray}
\end{widetext}
The first two terms in Eq.~(1) represent the electrons in the leads and in the impurities, respectively.
In these terms, $c^\dagger_{k_{L/R},\sigma}$ ($c_{k_{L/R},\sigma}$) creates (annihilates) an
electron with momentum $k_{L/R}$ and spin $\sigma$ in the left/right lead, and
$d^\dagger_{L/R\sigma}$ ($d_{L/R\sigma}$) creates (annihilates) an electron with spin $\sigma$ in the left/right impurity.
$\epsilon_{k_{L/R}}=\epsilon_{k}+\mu_{L/R}=\epsilon_{k}\pm\frac{eV_{sd}}{2}$ are the energies in the leads, while $\epsilon _{\alpha\sigma}$ are the bare energies at each impurity.
The third term describes the coupling between each impurity and its corresponding lead, and
determines the coupling strength
$\Delta_{L,R}(\epsilon)=\pi V_0^2\sum_{k_{\alpha\in \{ L,R \}}} \delta(\epsilon-\epsilon_{k_{\alpha}})$
(we neglect the $k$ dependency
of the tunneling matrix element for simplicity).
Each lead is described by a parabolic density of states (energy bandwidth $W=2D$) centered at the chemical potential,
such that we can define the function
\begin{widetext}
\begin{displaymath}
\Delta_\alpha(\epsilon)=\pi V_0^2\sum_{k_\alpha}\delta(\epsilon-\epsilon_{k_\alpha})= \left\{ \begin{array}{ll}
\Delta_0[1-(\frac{\epsilon-\mu_\alpha}{D})^2] &\textrm{if $-D\leq \epsilon-\mu_\alpha\leq D$,}\\
0 &\textrm{otherwise.}
\end{array}\right .
\end{displaymath}
\end{widetext}
The fourth term describes inter-impurity coupling.
In the absence of such coupling, this Hamiltonian describes two independent Anderson impurities
each of them coupled to different Fermi seas.
In the limit of $U_L,U_R\rightarrow\infty$
we can write the Hamiltonian
(1) in terms of auxiliary pseudo-fermions and slave boson (SB) operators plus constraints:
\begin{widetext}
\begin{eqnarray}
H&=&\sum _{ k_{\alpha\in \{ L,R \}}, \sigma} \epsilon _{k_\alpha} c^\dagger_{k_\alpha,\sigma}
c_{k_\alpha,\sigma}
+ \sum _{\alpha\in \{ L,R \},\sigma} \epsilon _{\alpha\sigma} f^\dagger_{\alpha\sigma}
f_{\alpha\sigma}
+\frac{t_I}{N}\sum _\sigma(f^\dagger_{L\sigma}b_L b^\dagger_R f_{R\sigma}
+f^\dagger_{R\sigma}b_R b^\dagger_L f_{L\sigma})\nonumber\\
&+&\frac{V_{sd}}{\sqrt N}\sum _{k_{\alpha\in \{ L,R \}},\sigma }(c^\dagger_{k_\alpha, \sigma} b^\dagger_\alpha
f_{\alpha\sigma}
+ f^\dagger_{\alpha\sigma} b_\alpha c_{k_\alpha, \sigma}).
\end{eqnarray}
\end{widetext}
In the slave boson representation, the annihilation operator for electrons at the impurity sites,
$d_{\alpha\sigma}$ is decomposed into the SB operator $b^\dagger_\alpha$
which creates an empty state and a pseudo fermion operator $f_{\alpha\sigma}$ which annihilates
the singly occupied state with spin $\sigma$ in the impurity $\alpha$:
$d_{\alpha\sigma}\rightarrow b^\dagger_\alpha f_{\alpha\sigma}$
($d^\dagger_{\alpha\sigma}\rightarrow f^\dagger_{\alpha\sigma}b_\alpha$).
Note that we have scaled the hopping parameters $V_0=\frac{V_{sd}}{\sqrt N}$ and $V_I=\frac{t_I}{N}$, $N$ being the degeneracy
of the level on each impurity. This scaling is done in such a way that the parameters
$\Gamma=N\Delta_0$
and $\Delta_0/V_I=N\Delta_0/t_I=\Gamma/t_I$ appearing in the expression of
the Kondo temperature have a well defined $N\rightarrow\infty$ limit,
namely there is a well defined $1/N$ expansion of the physical quantities.
At the end of the calculation, the physical limit $N=2$ is, of course, taken.
Finally, the physical constraint is that we must work in a subspace of the Hilbert space
where the number of auxiliary particles (on each impurity) is one, namely:
\begin{eqnarray}
\hat{Q}_L&=&\sum_\sigma f^\dagger_{L\sigma}f_{L\sigma}+b^\dagger_L b_L=1,\nonumber\\
\hat{Q}_R&=&\sum_\sigma f^\dagger_{R\sigma}f_{R\sigma}+b^\dagger_R b_R=1.
\end{eqnarray}
These two constraints come from the physical condition that
each impurity has to be in one of the three states $|0\rangle$,  $|\uparrow\rangle$ or $|\downarrow\rangle$.

At this point, we have reduced the original problem described by the
Hamiltonian in Eq.~(1) to a problem of fermions and bosons
interacting through tunneling terms and subject to the constraints
in Eq.~(3). Properties of the physical electrons can be build up
from the Green's functions of the pseudo-fermions and slave bosons.
These Green's functions for the auxiliary fermions and bosons
constitute the basic  building blocks of the theory. Furthermore,
our aim is to study  the out-of-equilibrium properties of the
system; we need, then, a fully non-equilibrium description of the
dynamics of the Green's functions of these auxiliary particles. The
appropriate starting point is to derive equations-of-motion for the
time-ordered double-time Green's function of the auxiliary fermion
and boson fields on a complex contour. In order to do this we employ
the so-called non-crossing approximation (NCA)
\cite{SI-NCAneq1,SI-NCAneq2,SI-NCAneq3,SI-NCAneq31} generalized to a
two-impurity Anderson Hamiltonian \cite{SI-ram03}. Without entering
into much detail of the theory, we just mention that the boson
fields are treated as fluctuating operators such that both thermal
and charge fluctuations are included in a self-consistent manner to
order $O(\frac{1}{N})$. In particular, one has to derive
self-consistent equations-of-motion for the time-ordered double-time
Green's function (sub-indexes are omitted here):
\begin{eqnarray}
iG(t,t')&\equiv&\langle T_c f(t)f^\dagger(t')\rangle\nonumber\,,\\
iB(t,t')&\equiv&\langle T_c b(t)b^\dagger(t')\rangle,
\end{eqnarray}
or in terms of their analytic pieces:
\begin{eqnarray}
iG(t,t')&=&G^{>}(t,t')\theta(t-t')-G^{<}(t,t')\theta(t'-t)\,,\nonumber\\
iB(t,t')&=&B^{>}(t,t')\theta(t-t')+B^{<}(t,t')\theta(t'-t).
\end{eqnarray}

A rigorous and well established way to derive these
equations-of-motion was first introduced by Kadanoff and Baym
\cite{SI-Kadan}, and has been related to other non-equilibrium methods
(like the Keldysh method) by Langreth, see ref.~\onlinecite{SI-lan76} for a
review. In the paper, we just show numerical results of the coupled
set of integral NCA equations for our problem and refer the
interested reader to
refs.~\onlinecite{SI-NCAneq1,SI-NCAneq2,SI-NCAneq3,SI-NCAneq31,SI-ram03} for details. In
particular, the density of states is given by
\begin{equation}
\rho(\omega)=-\frac{1}{\pi}\sum_{\sigma}
\mathrm{Im}[A^{r}_{\sigma}(\omega)],
\end{equation}
where $A^{r}_{\sigma}(\varepsilon)$ is the Fourier transform of
the retarded Green's function
$A^{r}_{\sigma}(t)=
G_{\sigma}^{r}(t)B^{<}(-t)-G^{<}_{\sigma}(t) B^{a}(-t)$. Note that this decoupling neglects vertex corrections and, as a result, the NCA fails in describing the low-energy Fermi-liquid regime. Nevertheless, the NCA has proven to give reliable results even at temperatures well below the Kondo temperature (of the order of $T=10^{-2}~T_K$) \cite{SI-Kroha2}.
Following Meir and Wingreen in ref. \onlinecite{SI-Meir92a}, the current is given by
\begin{equation}
I_{\alpha\in \{ L,R \}}=-\frac{2e}{h}\sum_{\sigma}\int d\epsilon
\Gamma_\alpha(\epsilon)[2Im A^r_{\sigma}(\epsilon)f_\alpha(\epsilon)
+A^<_{\sigma}(\epsilon)],\nonumber\\
\end{equation}
with $A^<_{\sigma}(\epsilon)$ the Fourier transform of
$A^{<}_{\sigma}(t)=
iG_{\sigma}^{<}(t)[B^{r}(-t)-B^{a}(-t)]$ and $f_\alpha(\epsilon)=\frac{1}{1+e^{\frac{(\epsilon-\mu_\alpha)}{kT}}}$ the Fermi-Dirac function at each reservoir held at a chemical potential $\mu_\alpha$ such that the applied bias voltage is defined as $eV_{sd}=\mu_R-\mu_L$.

In practice, we self-consistently solve the NCA integral equations for each isolated Anderson impurity until good numerical convergence is reached. In a second self-consistent step, we obtain the self-energies coming from inter-impurity coupling \cite{SI-ram03}.

\vspace{0.5cm}

\noindent \textbf{Parameters used for the theoretical traces in Fig.~\ref{Fig:FigTBdep}}

\noindent The theoretical traces in Fig.~\ref{Fig:FigTBdep} of the
main text are calculated as follows and for the following
parameters. The inset of Fig.~\ref{Fig:FigTBdep}a presents traces of
$dI/dV_{sd}$ calculated with the theory above here.
The numerics are performed by discretizing the Fourier space in a finite mesh of size $N_\omega = 2^{18}$ with cutoff $D = 20$.
The asymmetry primarily arises from taking unequal coupling strengths $\Gamma_L$ and $\Gamma_R$ between the impurities and their respective reservoirs.
Further reference to a value for $\Gamma$ assumes the relation $\Gamma = \frac{\Gamma_L+\Gamma_R}{2}$.
The traces in the inset of Fig.~\ref{Fig:FigTBdep}a are for $\frac{\Gamma_L}{\Gamma_R}=1.5$, and inter-impurity coupling $t_i=1.2 \: \Gamma$.
For the presentation in the main text all energy scales are with respect to the highest Kondo temperature $T_K$ of the two-impurities. We define it by using the following estimate for $T_K$ of each impurity,
\begin{equation}
T_K=D\sqrt{\frac{1}{\pi|\varepsilon_i|}} \cdot
\exp{\left( \frac{-\pi|\varepsilon_i|}{\Gamma_i} \right)}
,\nonumber\\
\end{equation}
where $\varepsilon_i$ is the energy of the Anderson impurity level $i=L,R$, and  $\Gamma_i$ should be filled in as $\Gamma_L$ or $\Gamma_R$ for the interaction of impurity $L,R$ with its respective reservoir.
(Note that for the single-impurity case with coupling to two reservoirs on two sides, one must use $\Gamma_i = \Gamma_L+\Gamma_R$, which yields a Kondo temperature that is significantly higher than the values for the two-impurity case.)
The bias voltage $V_{sd}$ is thus expressed in units of $k_B T_K/e$
($k_B$ is Boltzmann's constant).
The traces are for increasing temperature from $T=0.76 \: T_K$ (black
trace) to $T=1.33 \: T_K$ (red trace), see also the colored dots in the
inset of Fig.~\ref{Fig:FigTBdep}b. The inset Fig.~\ref{Fig:FigTBdep}b
is derived from these traces in the same manner as for the
experimental traces, simply by extracting the conductance between the
two peaks. We used here for all traces the conductance level at $V_{sd} = -0.88 \: k_B T_K/e$.

For calculations of magnetic field dependence, as presented in the inset of Fig.~\ref{Fig:FigTBdep}c, we simply added a Zeeman energy term for each of the localized spins, namely $\pm\Delta_Z/2$. The traces are for (top to bottom) Zeeman energies of $\Delta_Z=0$ to $\Delta_Z=11 \: k_B T_K$. It is well known that NCA (since it is a high-$N$ method) shows unphysical zero bias features at finite magnetic fields at low temperatures. Indeed, running the NCA for the single-impurity case reveals such a spurious anomaly. However, as one increases the temperature this feature goes away much faster than the physical features coming from Zeeman-split Kondo peaks. In order to avoid such spurious features, while obtaining good numerical convergence, we had to work with a better mesh resolution ($N_\omega = 2^{20}$) and more stringent convergence parameters (as a criterium for good convergence, the iteration stops when the relative error between successive occupations, as calculated from lesser Green's functions, in the iteration loop is less than $10^{-6}$).  Also, we use slightly different parameters (a bit larger inter-impurity coupling $t_i=1.3 \: \Gamma$ and a bit smaller asymmetry $\frac{\Gamma_L}{\Gamma_R}=\frac{1.1}{0.9}$) in order to resolve the Zeeman-split peaks without the need of going to lower temperatures that might be on the verge of reliability. This explains the slightly higher conductance values in Fig.~\ref{Fig:FigTBdep}c as compared to Fig.~\ref{Fig:FigTBdep}a (and it is known that this method can yield $dI/dV$ values that are slightly too high on the scale of $2e^2/h$ \cite{SI-NCAneq2}). Hence, these plots at finite magnetic fields are meant to show qualitative agreement. Quantitative agreement in such $dI/dV$ calculations at finite magnetic field and temperature is beyond this technique (or any other technique that we know of).

\newpage

\section{SDFT simulations}\label{Sec:SOMonSDFT}

\noindent \textbf{Summary of approach and main findings}

\begin{figure}[h!]
\centering
\includegraphics[height=12cm]{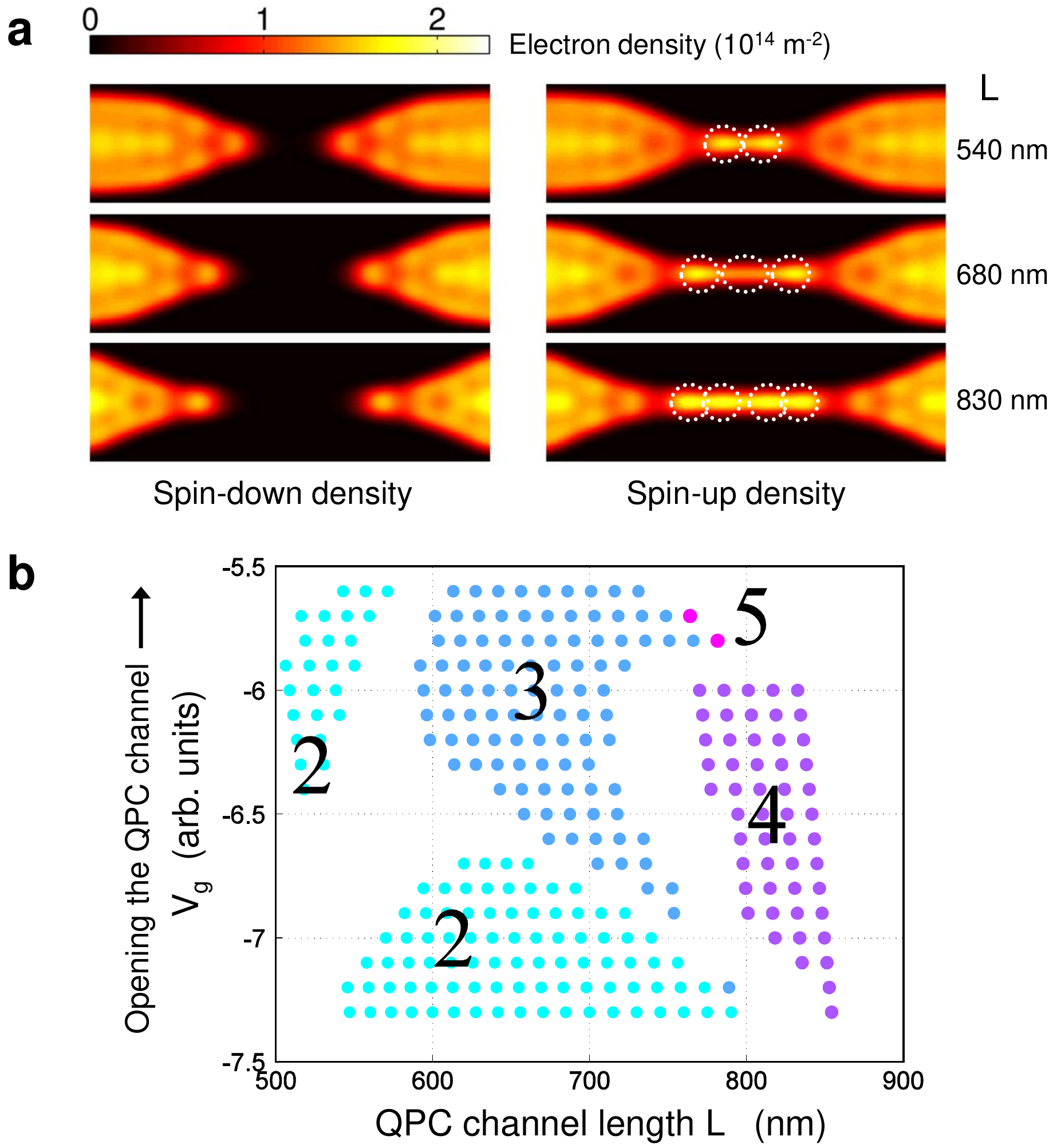}
\caption{\textbf{Results of Spin-Density-Functional-Theory calculations (SDFT)}.
\textbf{a}, Spin-polarised electron density in QPC channels (top view on 2DEG plane) for spin-up (right panels) and spin-down (left panels) at constant gate voltage $V_g=-6$ (arb.~units, giving a conductance just below $1 \cdot (2e^2/h)$) for three different lengths (540~nm, 680~nm and 830~nm). The number of ELSs (marked with white dotted circles) inside the QPC channel is 2, 3 and 4 respectively.
The number of ELSs increases by one each time when the QPC length increases over a range that allows for one additional period of the Friedel oscillation in the QPC channel (the Fermi wavelength in the plane 2DEG area was for these calculations 150~nm). The color scale extends from 0 (black) to $2\cdot 10^{14}~{\rm m^{-2}}$ (yellow).  \textbf{b}, The number of ELSs inside the QPC as a function of gate voltage $V_g$ (parameter for opening the QPC channel) and QPC channel length.}
\label{Fig:FigSDFT}
\end{figure}

\noindent To support the picture for the many-body effects that was presented in the main text we present numerical SDFT results for QPCs, generalizing earlier calculations \cite{SI-Hirose,SI-Rejec2006}, to allow for QPCs of changing length. The length of the channel was determined by a gate of variable length. Opening the QPC is controlled with a single gate-voltage parameter $V_g$.
We studied whether the SDFT yields localised states with about one electron of charge as the state with lowest energy. For the true ground state, these states should have spin-singlet character, but this cannot be addressed with SDFT
in the present approximation.
SDFT can still be used for checking whether a spin-polarised solution (in an arbitrary direction) has a lower energy than an unpolarised solution.
Since the polarisation direction is arbitrary, the SDFT then admits two degenerate ground states and the true ground state is a superposition of these two cases with spins in a singlet state. The electron density, though, is the same for the two degenerate SDFT ground states, and thus also for their linear superposition.
To allow for a spin-polarised solution, we started the iterative procedure with a finite magnetic field in arbitrary direction that breaks the spin-symmetry, which was turned off in subsequent iterations. For most of the values of gate voltage below the first plateau, the polarised solution had lower energy than the unpolarised solution (Fig.~\ref{DE}). An example of such SDFT results is presented in the Fig.~\ref{Fig:FigSDFT}
(but similar results are found for a wide range of parameters).
Figure~\ref{Fig:FigSDFT}a
displays the electron density for one of the ground states, for both spins, with increasing QPC length, demonstrating an increasing number of ELSs, denoted by circles (the number of ELSs was determined from the total density in the QPC, by fitting the density to a sum of two-dimensional Gaussians, each of total density $e$). Figure~\ref{Fig:FigSDFT}b
depicts the number of ELSs in the gate voltage-length plane, demonstrating the changing number of ELSs with these two parameters. This supports the picture we have presented in the main text: The number of ELSs increases by one each time when the QPC length increases over a range that allows for one additional period of the Friedel oscillation in the QPC channel.

\newpage



\newpage

\noindent \textbf{SDFT approach and methods}

\begin{figure}[ht]
\centering
\includegraphics[clip,width=0.7\hsize ]{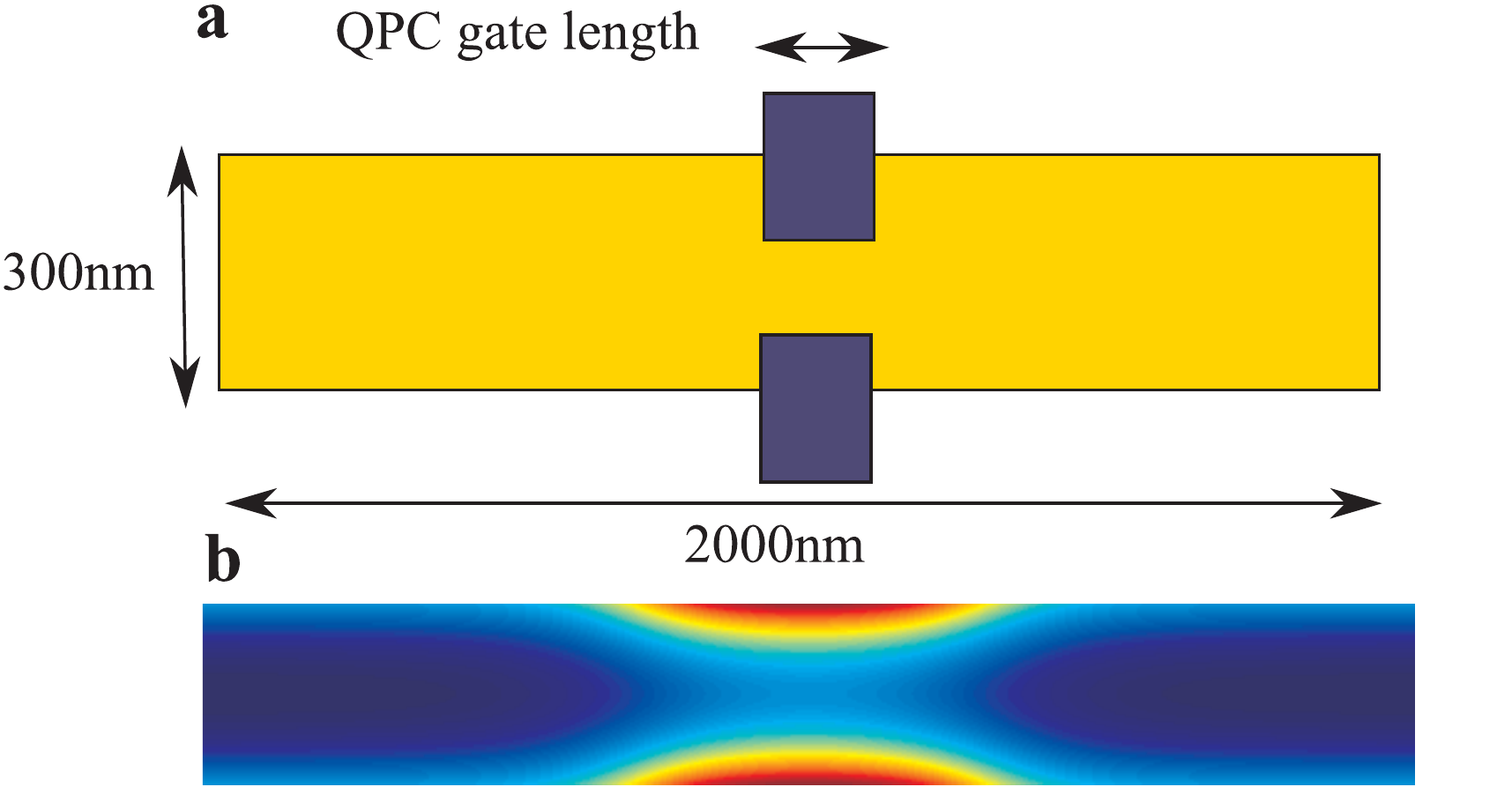}
\caption[Simulation Box]{\textbf{a}, Schematic of the simulation box, the blue rectangles represent the gate electrodes above the 2DEG. \textbf{b}, Example of the external saddle-point and wire potential for a QPC (blue is lowest, red is highest potential).}
\label{dftbox}
\end{figure}

\noindent The SDFT simulations were carried out in a rectangular box (representing a piece of 2DEG), shown in Fig.~\ref{dftbox}, with along the $x$-axis (horizontal in Fig.~\ref{dftbox}) periodic boundary conditions. The external potential is composed of a harmonic part $\frac{1}{2}\omega^2_y y^2$ that represents the wire and the QPC potential which is calculated by placing two negatively charged gate electrodes at height $z_0=100~{\rm nm}$ above the 2DEG in the middle, we are using the Yukawa potential
\be{1nr}
v(\mathbf{x_0},\mathbf{y_0})=\int \rho_g\frac{e^{- |\mathbf{r-r_0}|/\gamma}}{|\mathbf{r-r_0}|} d\mathbf{r}
\ee
where $\mathbf{r_0}=\sqrt{x_0^2+y_0^2+z^2_0}$, $\gamma=110~{\rm nm}$ and the integration is over the two-dimensional electrodes. The charge density of the electrodes $\rho_g$, is a function of the gate voltage on the electrodes $V_g$. In our simulation we set $\rho_g$, nevertheless, in order to present the results as in the experimental results, we regard it as gate voltage $V_g$ with arbitrary units.

For the Hartree term we use the appropriate two-dimensional system with one-dimensional periodic boundary conditions \cite{SI-Castro}. In addition, we add a positive image charge plane at height $100~{\rm nm}$ above the 2DEG as the contribution from the donor layer \cite{SI-Hirose}.

For the exchange and correlation functionals we use the local-density approximation, for the exchange we use slater exchange \cite{SI-exDirac} and the correlation functional is taken from quantum Monte-Carlo simulations of uniform electron gas \cite{SI-Attaccalite}. The total number of electrons is $N=108$ and the temperature is $300~{\rm mK}$. Though we have repeated some of the calculations with temperatures down to $60~{\rm mK}$, with very little change in the results. We used the Octopus code \cite{SI-octopus} for solving the equations. For all the simulations the electron effective mass $m_e^*=0.067m_e$ and the dielectric constant $\kappa=12.9$. The actual 2DEG electron density is taken slightly lower than in the experiments to keep the computational time of a simulation at a reasonable level, but we work in a regime where we capture the relevant physics. As a result, the relevant length scales (which are relative to the Fermi wavelength) are for the simulations also slightly longer than the experimental values.

The simulation steps are as follows:

\begin{itemize}
 \item Set an external potential for a given QPC gate length (parameterized by giving the gate electrodes a length $L$ along the channel) and given gate voltage of the electrodes. In this simulation, opening the QPC is thus controlled with a single gate-voltage parameter $V_g$ (in arbitrary units).
 \item Find the unpolarized ground state of the system by solving self-consistently the Kohn-Sham equations.
 \item Then we polarize the solution by applying a magnetic field perpendicular to the sample (only the Zeeman term) for a few iterations, turning it off and letting the system flow to its ground state again (in this procedure we are basically giving an educated initial guess for the density). In principle the field can be applied in any direction as it yields degenerate solutions.
 \item Repeat these steps for different QPC lengths and gate voltages.
\end{itemize}

\begin{figure}[h!]
\centering
\includegraphics[clip,width=0.66\hsize ]{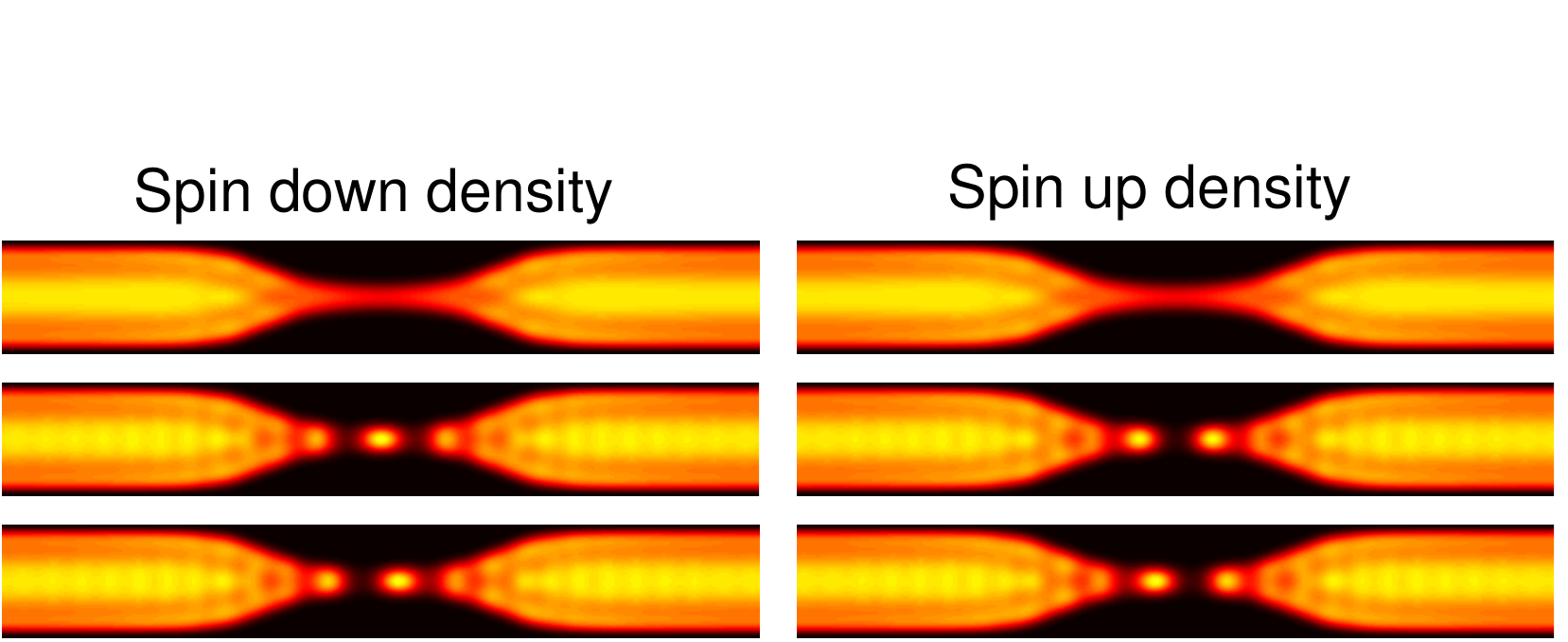}
\caption[Density example]{Spin-down (left) and spin-up (right) densities. Top row: unpolarized solution. Middle row: spatially-symmetric polarized solution. Bottom row: spatially-anti-symmetric polarized solution. The color scale extends from zero (black) to $2\cdot 10^{14}~{\rm m^{-2}}$ (yellow).}
\label{density1}
\end{figure}

\begin{figure}[h!]
\centering
\includegraphics[clip,width=0.49\hsize ]{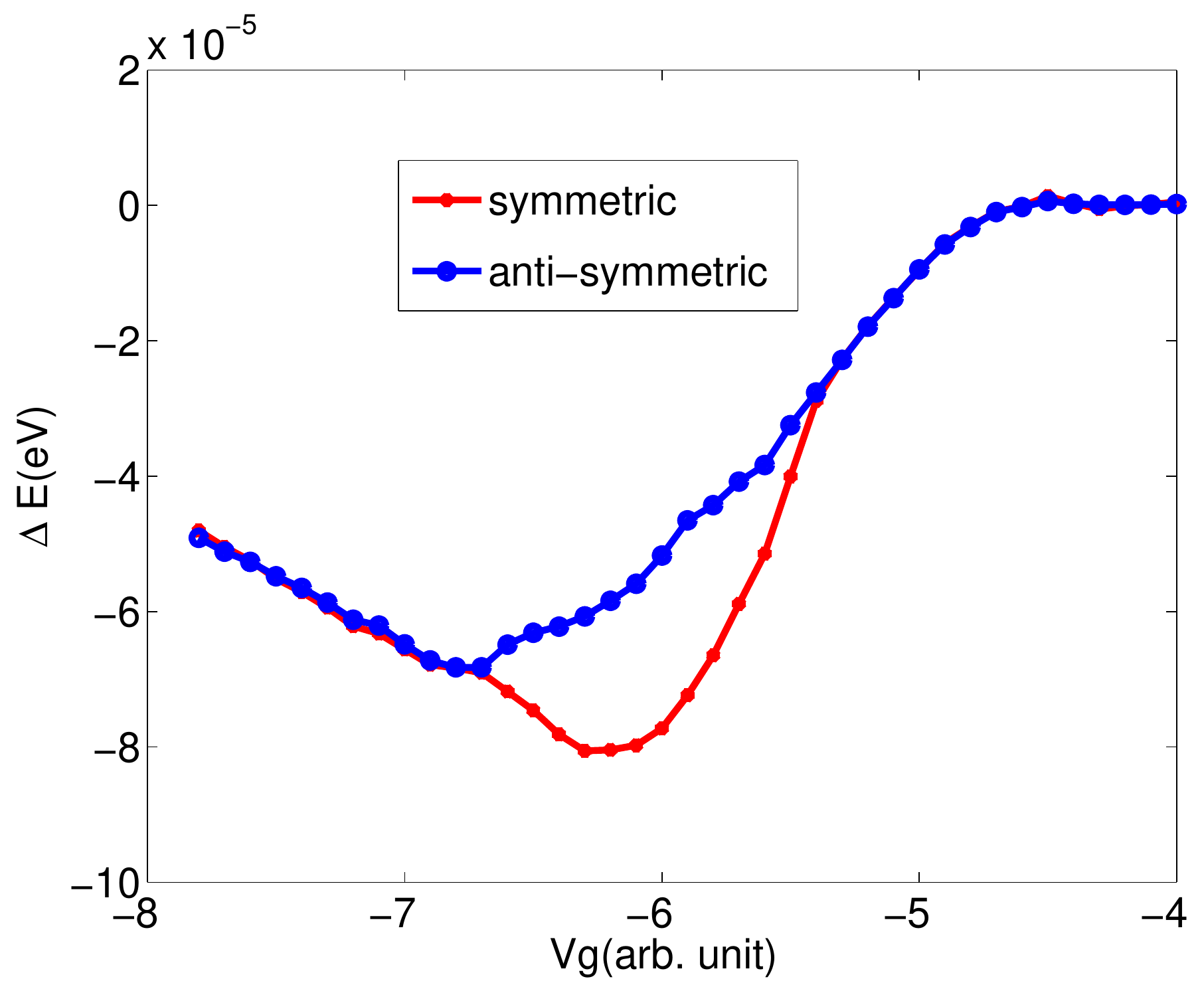}
\caption[Free energy as a function of gate voltage]{Free energies of the symmetric and anti-symmetric polarized solutions relative to the unpolarized solution as a function of gate voltage, for a QPC length of 680~nm.}
\label{DE}
\end{figure}

Figure~\ref{density1} shows two examples of the resulting densities, the left column is spin-down density while the right column in spin-up density. The top row presents the unpolarized solution, the middle row the spatially-symmetric polarized solution, and the bottom row the spatially-anti-symmetric polarized solution. The spatial symmetry of the magnetic field determines the symmetry of the solution.

Figure~\ref{DE} presents the Free energy difference $\Delta E$ between the polarized solutions relative to the unpolarized solution. If there exists a polarized solution it has a lower energy than the unpolarized solution. Moreover, the spatially-symmetric solution has a region $V_g=-[6.5-5.5]$ where it is the ground state of the system. In the following we will concentrate on this region which is below the first plateau.

\begin{figure}[h!]
\centering
\includegraphics[clip,width=0.49\hsize ]{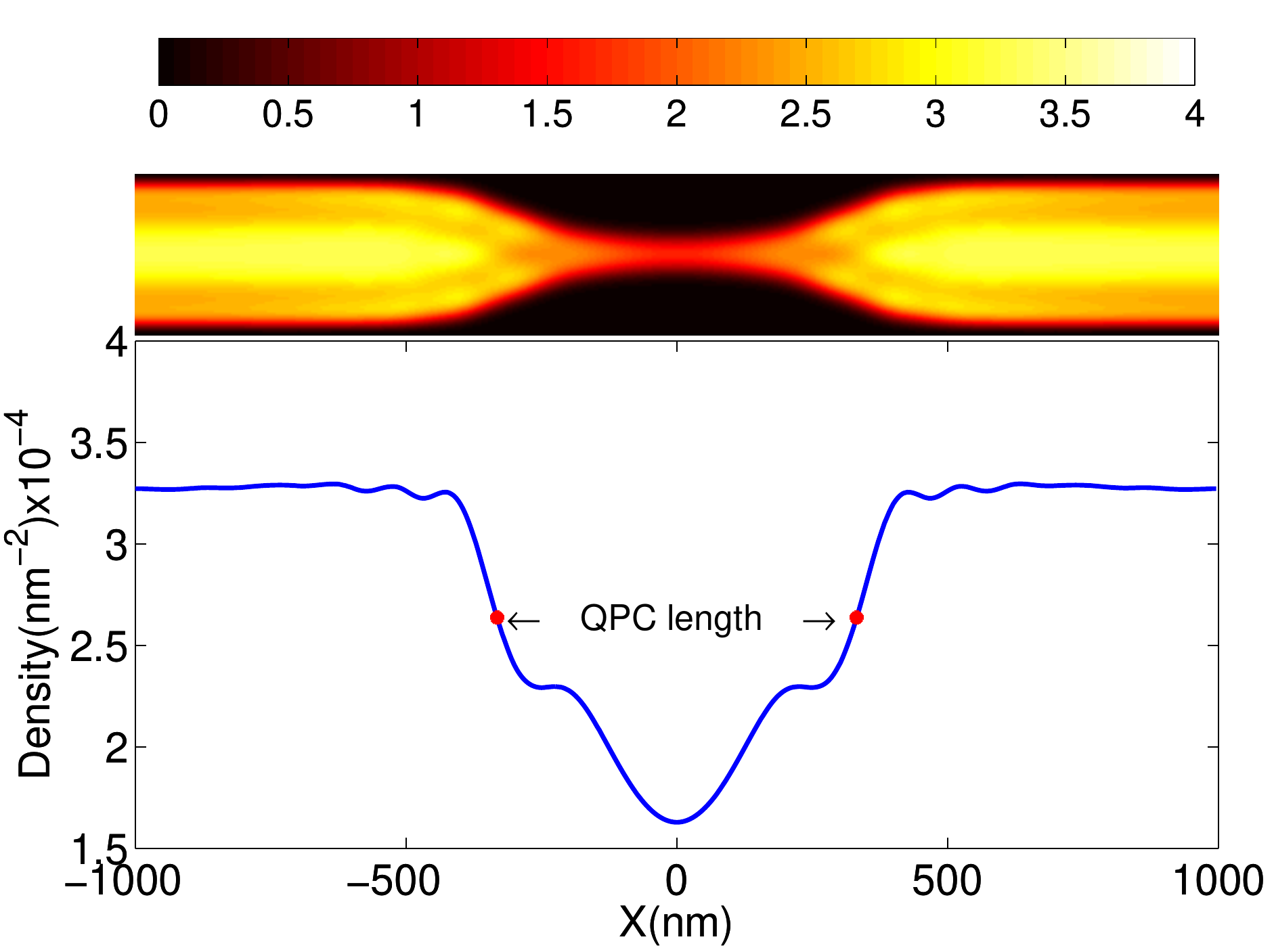}
\caption[QPC length calculation]{Determination of the effective length of the channel in the QPC saddle-point potential (typically shorter than the lithographic gate length $L$). The blue line is a cross section of the total density across the middle of the QPC channel. Above is the total electron density zero (black) to $3.5\cdot 10^{14}~{\rm m^{-2}}$ (yellow). The red dots mark the density at 80\% of the maximum density, and the distance between these points is used as the effective QPC channel length.}
\label{length_cal}
\end{figure}

To determine the number of emergent localized states (ELSs) that are in the effective QPC channel we cannot use that they appear in the region that is determined by the lithographic gate length $L$. Instead, the effective QPC channel appears as a saddle-point potential that has an effective length that is typically shorter than $L$. We used the following approach to define whether localized maxima in the charge density can be interpreted as an ELS within the effective QPC channel. We study the unpolarized solution, and define that the effective QPC channel is located between the two points at a density of 80\% of the maximum density, along a QPC cross section as depicted in Fig.~\ref{length_cal}.

\begin{figure}[h!]
\centering
\includegraphics[clip,width=0.49\hsize ]{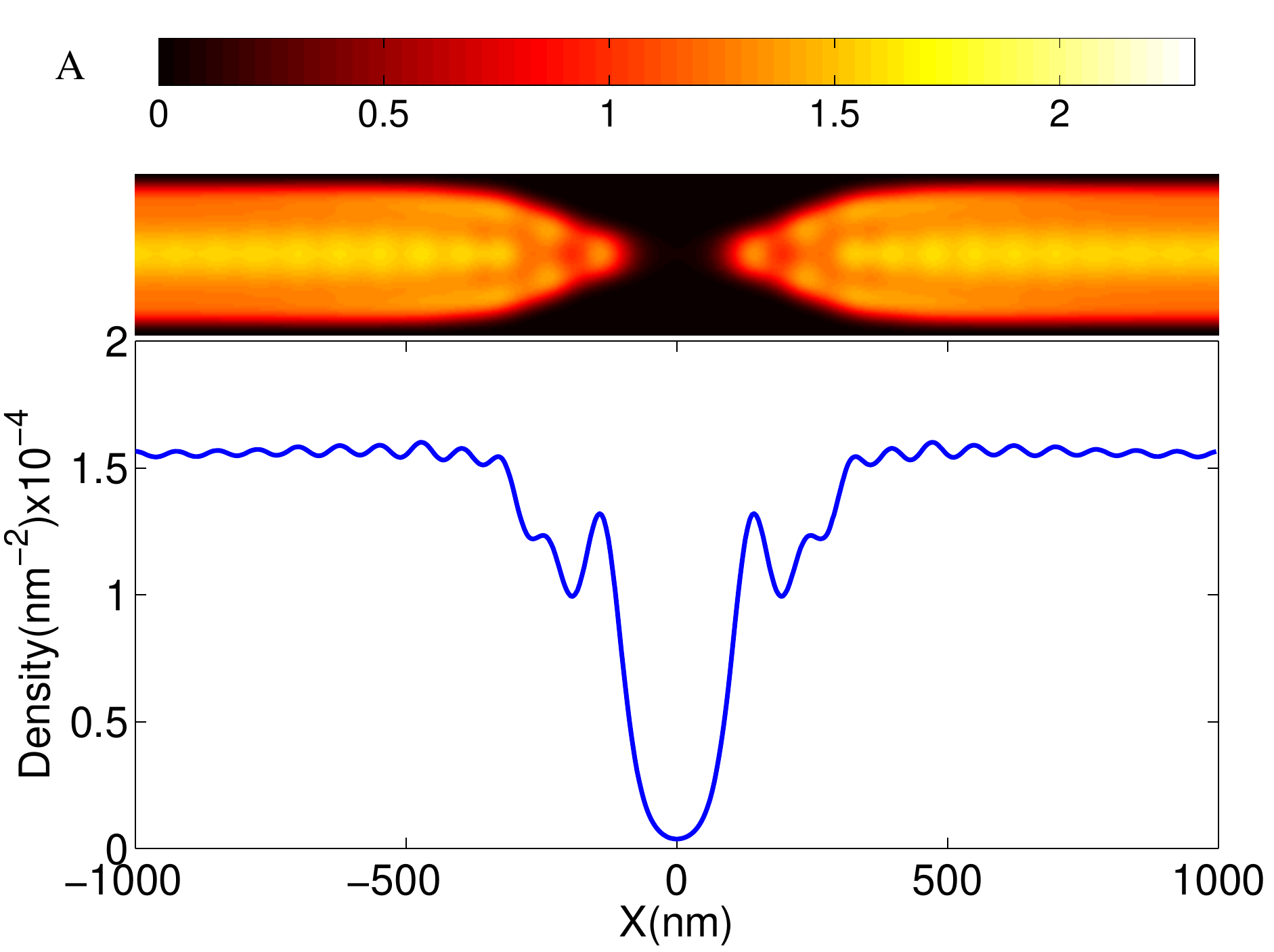}
\includegraphics[clip,width=0.49\hsize ]{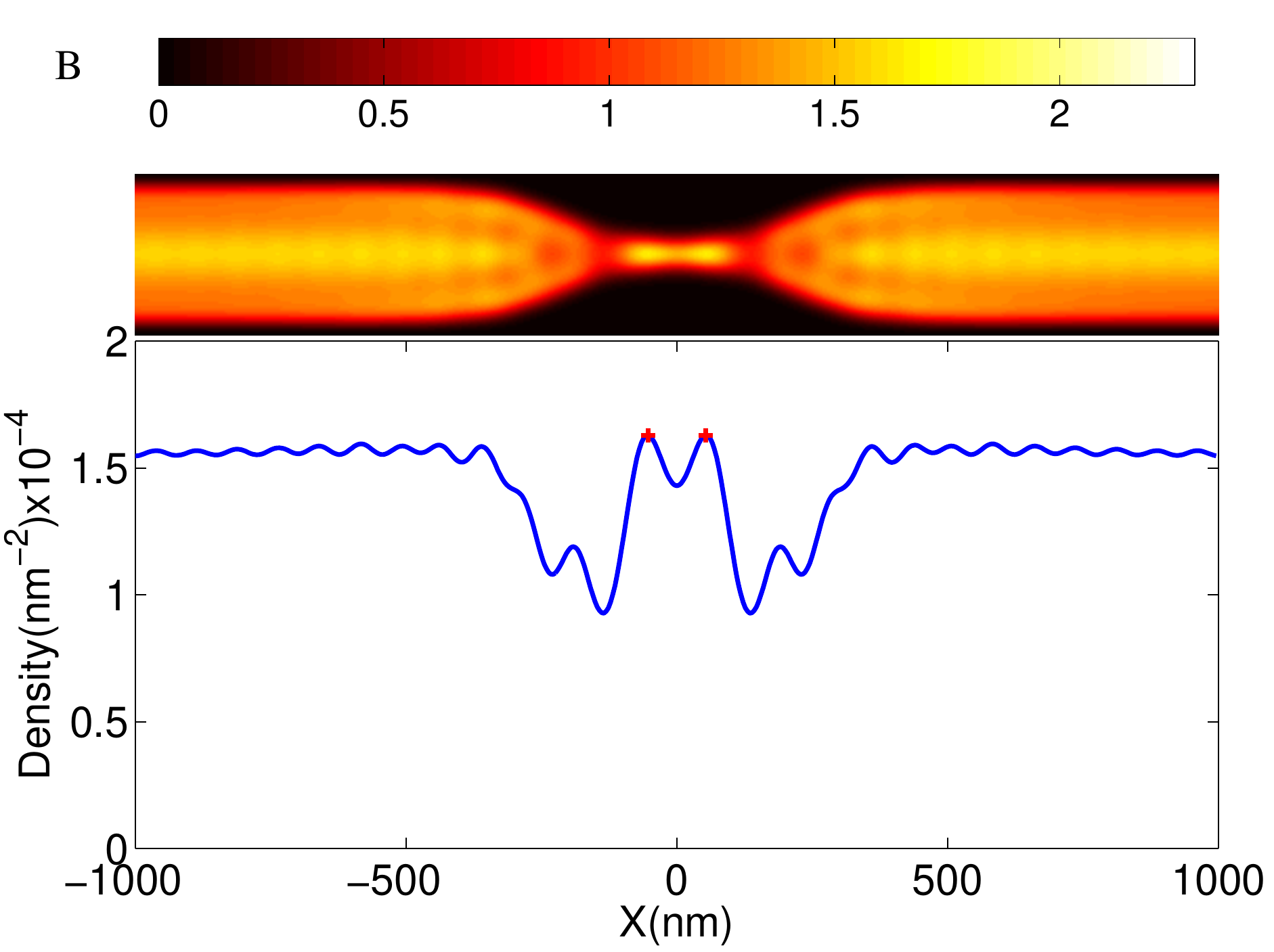}
\includegraphics[clip,width=0.49\hsize ]{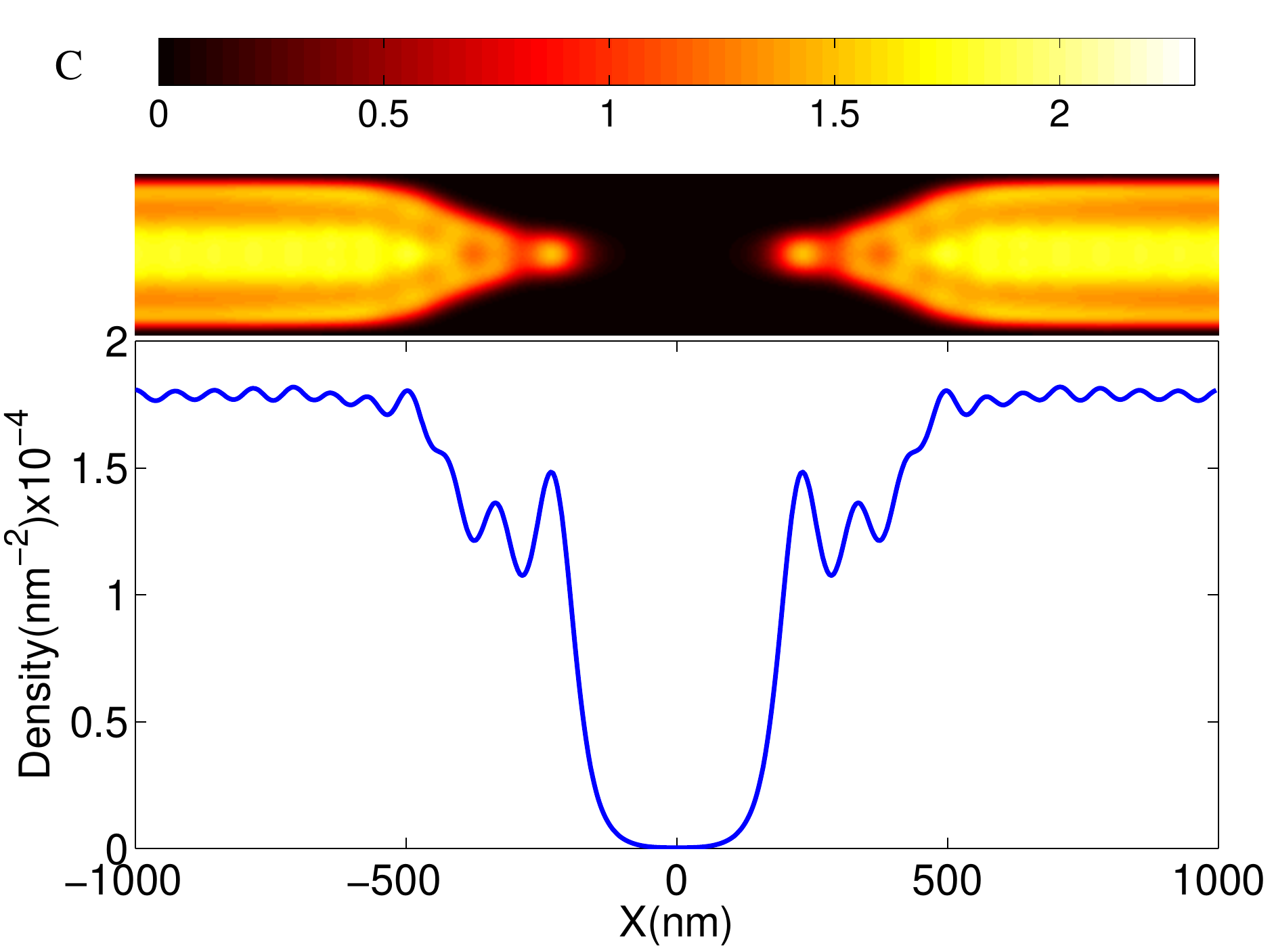}
\includegraphics[clip,width=0.49\hsize ]{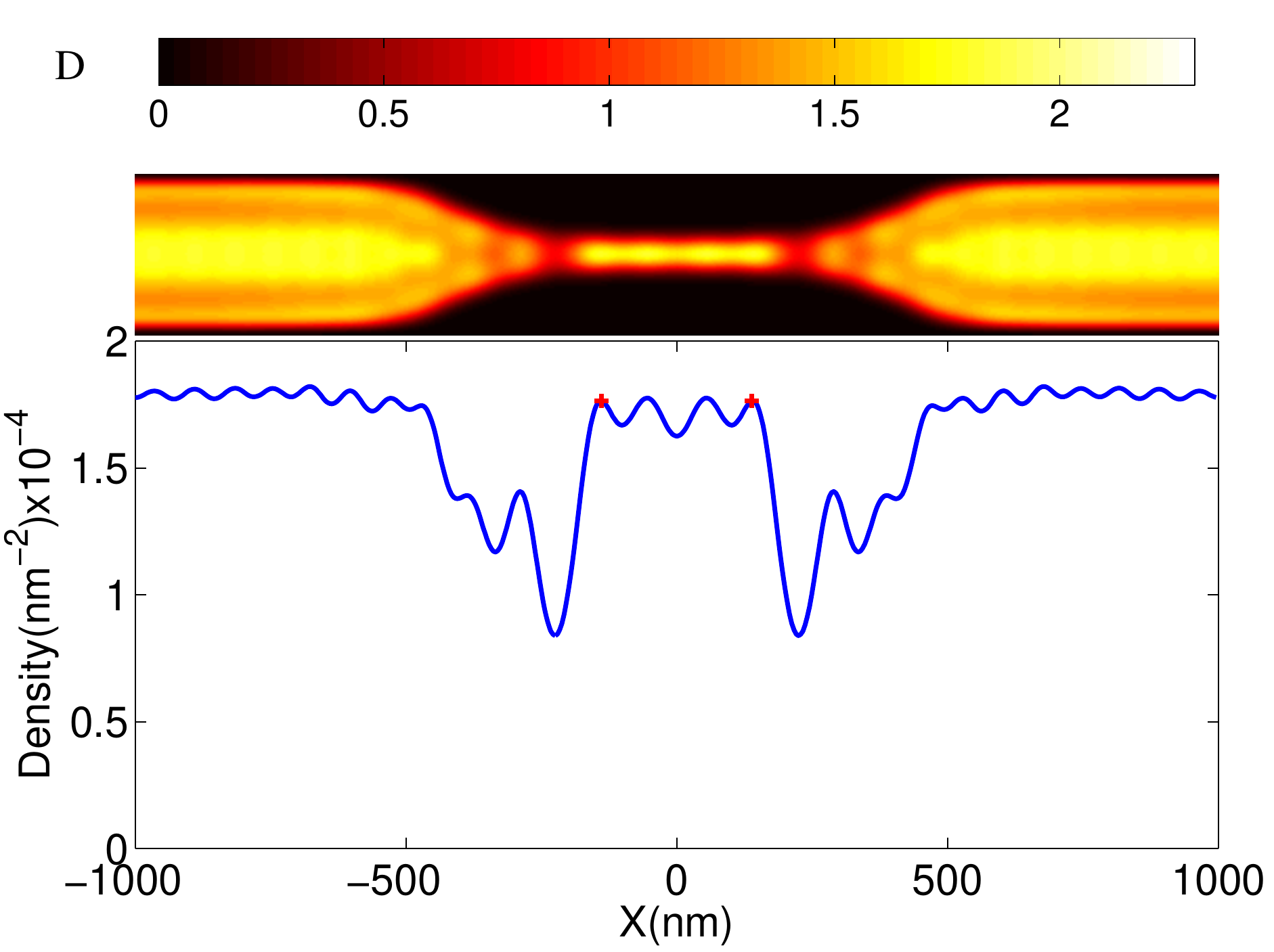}
\caption[Example for counting ELSs]{Example for counting ELSs - first method. \textbf{a},\textbf{b}, Electron density and its cross section for spin up (\textbf{a}) and spin down (\textbf{b}) at $V_g=-6$ and for $L=540~{\rm nm}$. The red crosses mark the Friedel oscillations that are counted as ELSs. Here we have a total of two ELSs in the QPC, each one has a density very close to one electron density.
\textbf{c},\textbf{d},Cross section of spin-up (\textbf{c}) and spin-down (\textbf{d}) densities at $V_g=-6$ and for $L=830~{\rm nm}$. In this case there are four ELSs, as can be seen from the cross section.}
\label{imp_cal}
\end{figure}

\begin{figure}[h!]
\centering
\includegraphics[clip,width=0.37\hsize ]{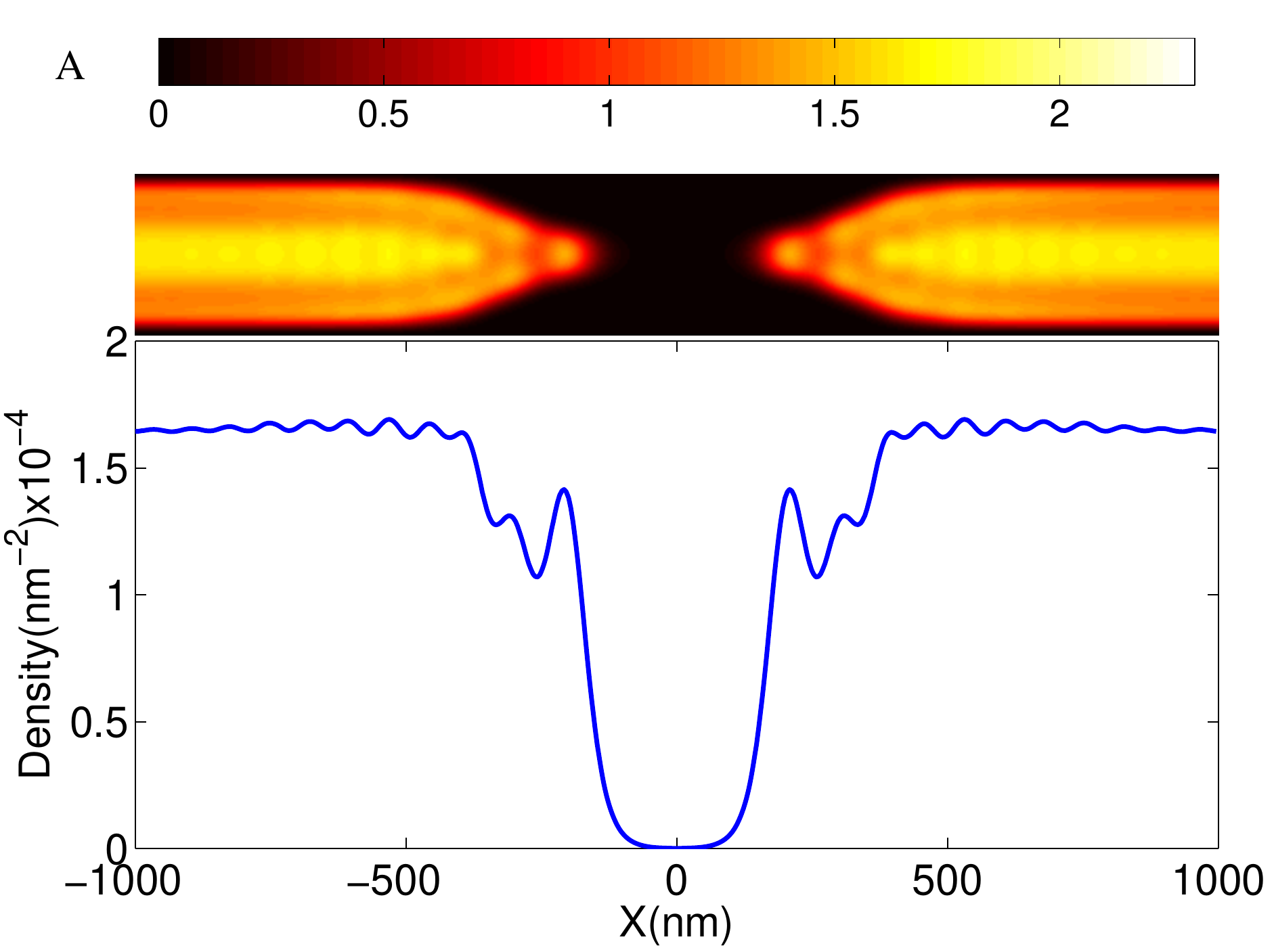}
\includegraphics[clip,width=0.37\hsize ]{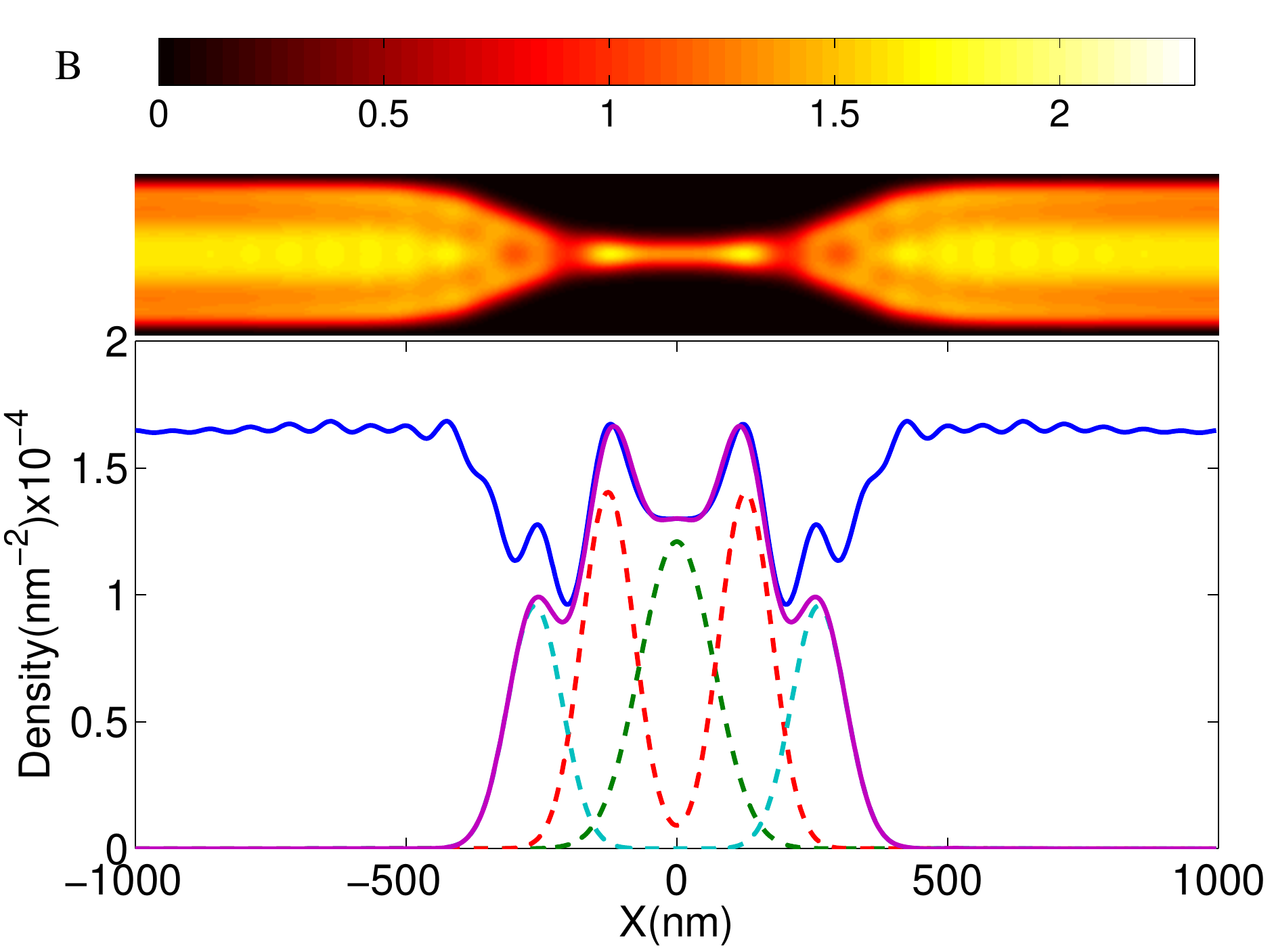}
\includegraphics[clip,width=0.16\hsize ]{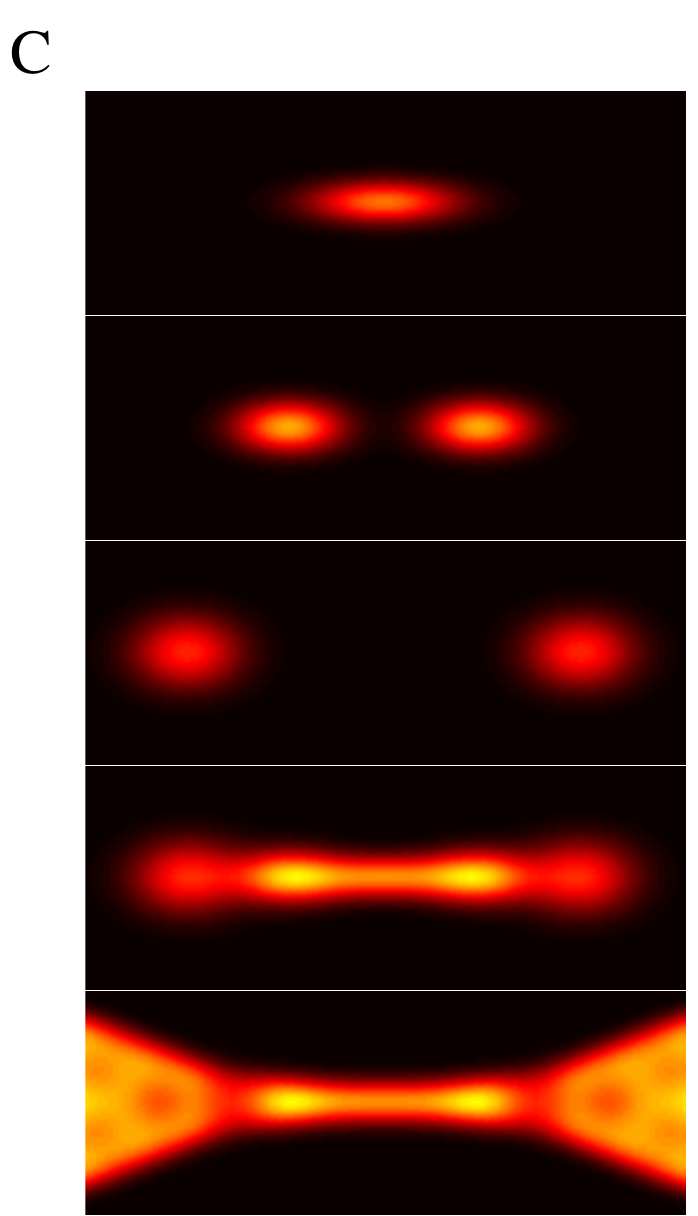}
\caption[Example for counting ELSs 2]{Example for counting ELSs - second method. \textbf{a},\textbf{b}, Electron density and its cross section for spin up (\textbf{a}) and spin down (\textbf{b}) at $V_g=-6$ and for $L=680~{\rm nm}$. Here the integration of the density inside the QPC gives a total of three electrons, though there are only two peaks in the cross section. The two dimensional density can be fitted to a sum of 5 unit weight ($e$) Gaussians (all 5 Gaussians have weight $e$ when accounting for their transverse dimension). Three of these Gaussian are within the effective QPC channel (Fig.~\ref{length_cal}).
\textbf{c}, We show how the five Gaussians give the corresponding SDFT density. From top to bottom (and with reference to traces in panel \textbf{b}): Center Gaussian (corresponds to green dashed line in the cross section), two nearest Gaussians (red dashed line), two next nearest Gaussians (cyan dashed line), all five Gaussians together (purple line), SDFT density (blue line). Hence we conclude that there are three localized ELSs inside the QPC.}
\label{imp_cal2}
\end{figure}

As was discussed in the main text, the ELSs inside the QPC originate from Friedel oscillations and as the QPC gets longer the Friedel oscillations have more periods in the QPC channel. We use two complimentary procedures in order to determine the number of ELSs inside the QPC. First we look at the cross section of the density across the middle of the sample (this is done separately for spin-up density and spin-down density). The criteria for the definition of a peaked feature that represents an ELS is given by
\be{2nr}
n_{\sigma}(\mathbf{r}_{max})*P \geq n_{\sigma}(\mathbf{r}_{min}),
\ee
$n_{\sigma}$ is the density of spin $\sigma$, and $\mathbf{r}_{min}$ is the closest minimum toward the outside of the QPC. This procedure is shown in Fig.~\ref{imp_cal} for $P=0.32$. As can be seen in Fig.~\ref{imp_cal}d, this procedure may be problematic when the solution is ferromagnetic, and the peaks overlap significantly. In such cases we use a second procedure: An example is given in Fig.~\ref{imp_cal2}. The total two-dimensional density for the up spin (panel \textbf{b}) can be fit to a sum of five Gaussians (only three of them inside the actual QPC channel), each of the total unit weight ($e$), even though the cross section reveals only two peaks inside the QPC.

\newpage

\newpage

\newpage

\newpage

\newpage

\section{Earlier reports showing modulated 0.7 anomalies and double-peak ZBA${\rm {\bf s}}$}\label{Sec:SOMotherReports}

A remarkable aspect of our observations of double-peak ZBAs in many
different QPC devices is that such observations were to our
knowledge never explicitly reported before, despite about 20 years
of research into many-body effects in QPCs (with a few hundred
publications reporting on the topic). Similarly, our observations of
a periodic modulation of the 0.7~anomaly as a function of QPC
channel length is a phenomenon that -to our knowledge- has not been
reported before. It is therefore justified to ask why this is the
case, and to question whether our -seemingly unique- observations
result from an  irregularity that is unique to our devices. The main
text already reported that we observed double-peak ZBAs in many
conventional QPC$_{{\rm 2F}}$ devices from two different wafer
materials and in 8 different QPC$_{{\rm 6F}}$ devices (and for both cases for multiple cool-downs and for different gate settings during cool-down), which all together provides convincing evidence that the new phenomena are not due to a particular device irregularity.

We noticed in our studies on many QPC$_{{\rm 2F}}$ that a strong
expression of the many-body effects that we report here appears as
quantized conductance plateaus that are not flat, and a 0.7~anomaly
that appears as a peaked resonance instead of only a shoulder on the
step from $G=0$ to $1 \cdot (2e^2/h)$ (see for example also
Fig.~\ref{Fig:FigTdepLIN}a near $V_{g1}=-0.35~{\rm V}$). These two
signatures can then also show replicas at and just below the second
and third quantized conductance plateau. In such cases, the ZBA
often has double or triple-peak character with strong asymmetries
and smaller side peaks. Such QPC results were in fact observed long
before by our group, but such results (and further studies of the
devices) were discarded because a device imperfection was suspected.
We know that most groups in the field had the same practice (see for
example ref.~\onlinecite{SI-Cronenwett2001thesis}). Also, such imperfections
on quantized conductance traces were often removed by applying a
small (typically 25~mT) perpendicular magnetic field during the QPC
studies, or by measuring at about 600~mK instead of the lowest
available temperature (see also ref.~\onlinecite{SI-Cronenwett2001thesis},
not applied in any of our studies). Only our recent study on a large
number of QPC$_{{\rm 2F}}$ showed us that such strong deviations on
quantized conductance traces fall in fact on a regular trend from
weak to strong expression of the 0.7~anomaly and the ZBA.
Nevertheless, it remains difficult to investigate this
systematically since the appearance is very diverse, and for results
from one particular QPC$_{{\rm 2F}}$ it remains a challenge to rule
out a (partial) role for an actual material or device defect.

In the remainder of this Section we report that several signatures
of double-peak ZBAs and periodic modulation of the 0.7 anomalies can
in fact be recognized in the experimental data of earlier
publications (but they were never discussed or systematically
investigated). Below here we review the literature on this, first
for modulation of the 0.7~anomaly and subsequently for double-peak
ZBAs. For both phenomena it is also illustrative to inspect all
experimental data in the extensive review ref.~\onlinecite{SI-Micolich2011}.

\vspace{0.5cm}
\noindent \textbf{Reports showing modulated 0.7~anomalies}

\noindent Several publications report a very regular modulation of
the 0.7~anomaly over about 1 period as a function of back(top)-gate
voltage in experiments that have a large-area back(top) gate in
addition to the split-gate structure of the QPC. A deviation from
our experiment is that such a back-gate modulates at the same time
the QPC saddle-point potential and the electron density of the
reservoirs. A regular modulation of the 0.7~anomaly over about 1
period with a fixed density for the reservoirs was also reported
before, in experiments with split-gate QPCs that have an additional
side gate or narrow top gate. Examples are ref.~\onlinecite{SI-Pyshkin2000}
(Fig.~2), ref.~\onlinecite{SI-Cronenwett2001thesis} (Fig.~6-17, left panel
on QPC2 on p.~105), ref.~\onlinecite{SI-Reilly2001} (Figs.~2, 3a and 3b),
ref.~\onlinecite{SI-Hew2008} (Fig.~2), and ref.~\onlinecite{SI-Wu2012} (Fig.~1). The
latter three references show in fact modulation over about 1.5
period, with an anomaly at a conductance level of about $0.9 \cdot
(2e^2/h)$ that disappears while a new one appears at about $0.6
\cdot (2e^2/h)$ (as in our data). Similar modulation is observed in ref.~\onlinecite{SI-Zhang2013} (Fig.~1).

\vspace{0.5cm}
\noindent \textbf{Reports showing double-peak ZBAs}

\noindent We also found several publications that report double-peak
ZBAs for data taken at zero magnetic field. An example that looks
much like our data is in ref.~\onlinecite{SI-Cronenwett2001thesis},
Fig.~6-19a on p.~107. Other examples can be found in
ref.~\onlinecite{SI-Patel1991} (Fig.~4) and ref.~\onlinecite{SI-Chen2009} (Figs.~1a,
1c, 3b and 3d). A very recent publication by Zhang \textit{et al.} (ref.~\onlinecite{SI-Zhang2013}) presents a few examples of double- and triple-peak ZBAs (Fig.~3).

\newpage

\section{Remote imperfections and reducing charge noise in QPC devices}\label{Sec:SOMremoteImp}

Very weak static fluctuations on the QPC saddle-point potentials are to be expected.
They can, for example, result from the small device-to-device
varation that is inherent to the nanofabrication process. In
addition, our devices have a doping layer at about 40~nm distance
from the QPC channel. In this doping layer ionized doping centers
are at random positions with an average inter-dopant distance of
about 10 nm.

Such weak static fluctuations on the QPC saddle-point potentials can have a strong influence on the expression of many-body effects.
For example, the parameter $\Gamma$ that was introduced in the main text for the coupling strength between a localized electron and a neighboring reservoir depends exponentially on weak potential fluctuations that are present between the localized state and the reservoir.

For most of our measurements, we stabilized the dynamical character of the (non)ionized doping centers by cooling down with a small positive voltage on the gates
\cite{SI-Buizert2008prl} (see also Methods of the main text).

\newpage

\section{Defining and calibrating ${\bf L_{eff}}$ for length-tunable QPC${\rm {\bf s}}$}\label{Sec:SOMcalLeff}

As introduced in the main text, operating the QPC$_{{\rm 6F}}$ devices induces a smooth saddle-point potential (as in Fig.~\ref{Fig:FigLtuneQPC}b) for which the length along the transport direction is controlled by operating at a particular ratio $V_{g2}/V_{g1}$. For such a smooth saddle-point potential it is not obvious what the value is of the channel length. We characterize this channel length with the parameter $L_{eff}$, which corresponds to the value of the lithographic length of a QPC$_{\rm 2F}$ type device (the length $L$ in Fig.~\ref{Fig:FigQPC2F}a) that gives effectively the same saddle-point potential.

\begin{figure}[h!]
\begin{center}
\includegraphics[width=8cm]{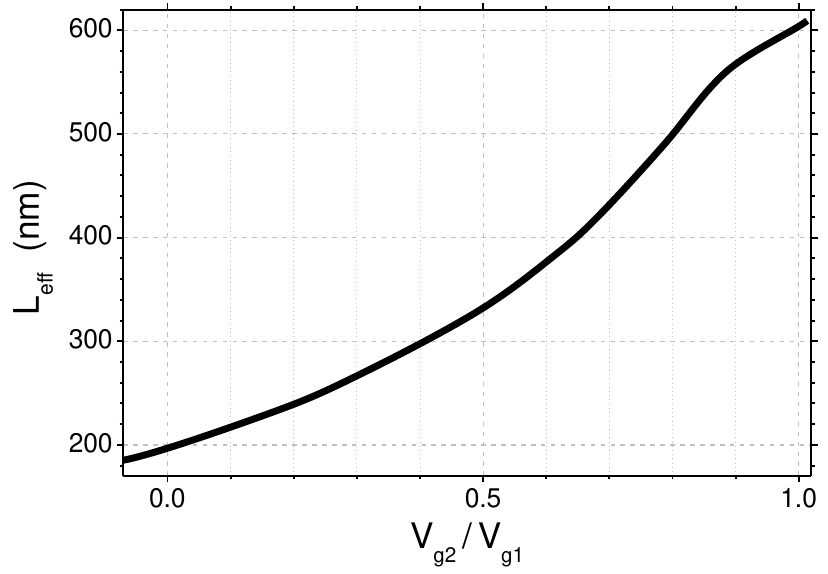}
\caption{Result of calibrating the dependence of $L_{eff}$ on $V_{g2}/V_{g1}$.}
\label{Fig:Lcalibrate}
\end{center}
\end{figure}

The results of calibrating the relation between $L_{eff}$ and $V_{g2}/V_{g1}$ are presented in Fig.~\ref{Fig:Lcalibrate}. A detailed account of this calibration can be found in a separate publication by our team \cite{SI-Iqbal2013}. This work also investigated whether there is significant structure on the saddle point potential from the narrow gaps (44~nm wide) between the three gate fingers on each side of the QPC channel. The results show that we operate the QPC$_{{\rm 6F}}$ under conditions far away from the point where such structure would become significant. The electron flow is 200~nm away from the gate electrodes such that effective electrostatic potential only reflects the gate geometry in a highly rounded manner.

Opening or closing the transport channel for QPC$_{{\rm 6F}}$ at a particular value for $L_{eff}$ requires co-sweeping of $V_{g1}$ and $V_{g2}$ at a fixed ratio $V_{g2}/V_{g1}$. Notably, in the case of cooling down with about +0.3~V bias on the gates (see Methods of the main text, and ref.~\onlinecite{SI-Buizert2008prl}), co-sweeping of $V_{g1}$ and $V_{g2}$ was carried out with respect to $V_{g1}=V_{g2}= + 0.3~{\rm V}$ instead of $V_{g1}=V_{g2}= 0~{\rm V}$ (for details see ref.~\onlinecite{SI-Iqbal2013}).

\newpage

\section{Friedel oscillations and Fermi wavelength in the QPC}\label{Sec:SOMfriedel}

The approximate form of a Friedel oscillation for screening of a charged scatter center at position $x=0$ (valid for positions not too close to $x=0$) is
\begin{equation}
\rho_{devi} \propto \frac{\sin(2k_F x + \phi)}{x^d},\nonumber
\label{Eq:Friedel}
\end{equation}
where $\rho_{devi}$ is the deviation in electron charge density, $d$
is the dimensionality of the Fermi liquid, $k_F$ is the Fermi
wavenumber, and $\phi$ is a phase parameter that depends on the
details of the scattering \cite{SI-Simion2005}. The factor 2 before
$k_F$ shows that the wavelength of Friedel oscillations is half the
Fermi wavelength.

The electron reservoirs in our experiment had a density $n_{2D}=1.6
\cdot10^{15}~{\rm m^{-2}}$, which corresponds to a Fermi wavelength
$\lambda_F=2\pi/k_F=62~{\rm nm}$. In the QPC channel the electron
kinetic energy is reduced. For an estimate we assume that the
kinetic energy ($E_F=5.7~{\rm meV}$ for the reservoir) gets reduced
in the QPC entries to a value of about 1~meV, which is estimated by
taking half the 1D subband energy spacing of our type of QPCs (value taken from our analysis in ref.~\onlinecite{SI-Koop2007}). This yields that $\lambda_F$ increases here to about
150~nm, which corresponds to about 75~nm for the Friedel oscillation
wavelength in the QPC channel. When increasing the length of the
QPC, the number of Friedel oscillations in the channel increases at
the same time (more or less symmetrically) in both entries of the
QPC. Thus, a length dependence that relies on the number of Friedel
oscillations in the channel should show a modulation with a
periodicity of about 150~nm, or a bit smaller ($\sim$100~nm) if the
effective Fermi wavelength in the QPC entries is still a bit closer
to the value for the reservoirs. Figure~\ref{Fig:FigZBAposAll} shows
that this is well in the range of the observed periodicity. The fact
that the modulation occurs faster around $L_{eff}=250~{\rm nm}$ than
around $L_{eff}=500~{\rm nm}$ is consistent with the fact that the
latter case corresponds to a longer shallow channel (less abrupt
saddle-point potential) where the Fermi wavelength is extended over a longer range.

\newpage

\section{Temperature dependence of linear conductance}\label{Sec:SOMdataTdepLIN}

\begin{figure}[h!]
\centering
\includegraphics[width=16cm]{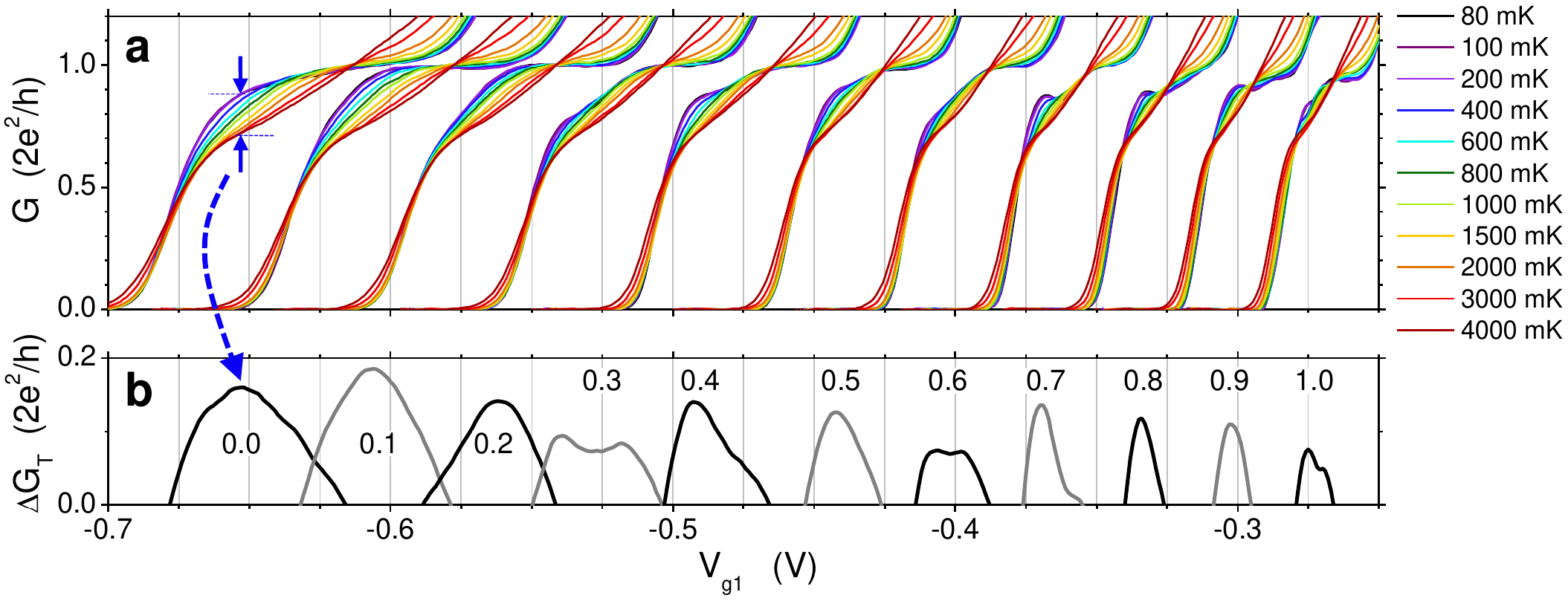}
\caption{Temperature dependence of the linear conductance.
\textbf{a}, Evolution of the linear conductance traces of Fig.~\ref{Fig:FigLtuneQPC}c as a function of temperature. Traces are displayed for gate ratios $V_{g2}/V_{g1}=0.0$, $0.1$, $0.2 \ldots 1.0$, as labeled in \textbf{b}.
\textbf{b}, The difference in linear conductance $\Delta G_{T}$ between the 80~mK and 4000~mK traces of panel \textbf{a}. These traces directly reflect the enhancement of the linear conductance around the 0.7~anomaly with decreasing temperature.}
\label{Fig:FigTdepLIN}
\end{figure}

Figure~\ref{Fig:FigTdepLIN} presents for the same device as used for Fig.~\ref{Fig:FigLtuneQPC}, \ref{Fig:FigZBAcharact} and \ref{Fig:FigTBdep} in the main text results for the temperature dependence of the 0.7 anomaly in linear conductance traces. At the
highest temperature (4000~mK), the linear conductance traces no
longer show quantized conductance plateaus (Fig.~\ref{Fig:FigTdepLIN}a)
and the only remaining feature is the 0.7 anomaly, which no longer
shows a modulation as a function of $L_{eff}$. For the following
discussion we focus on gate settings that give $G \approx 0.7 \cdot
(2e^2/h)$ at 4000~mK. At these points, the linear conductance
increases from 0.7 towards $1 \cdot (2e^2/h)$ when lowering the
temperature. Notably, the increase in conductance is for all
$L_{eff}$ for the largest part due to a growing height of the ZBA
(observed in the corresponding nonlinear conductance results). Also,
subtracting 4000~mK traces from 80~mK traces (defining the traces
$\Delta G_{T}$ in Fig.~\ref{Fig:FigTdepLIN}b) shows that the
conductance increase is largest around these points. However, some
$\Delta G_{T}$ traces show a suppression (for $V_{g2}/V_{g1}=0.3$,
0.6 and 1.0 in Fig.~\ref{Fig:FigTdepLIN}b) and these points coincide
with a strong 0.7 anomaly at the lowest temperatures and pronounced
double-peak character for the ZBA. That is, the $\Delta G_{T}$
curves show that the enhancement of $G$ due to many-body effects is
strongest where the linear conductance is about $0.7 \cdot (2e^2/h)$
at high temperatures, but that there is a range within each
$L_{eff}$ period where exactly at this point the strongest
expression of a new effect causes in fact a dip in $\Delta G_{T}$.
Further analysis shows that this coincides with the points where the
double-peak ZBA shows behavior that is characteristic for the
two-impurity Kondo model, and that the energy scale for the coupling
between the two spins in this model appears maximum at this point.

\newpage

\newpage

\section{Magnetic-field dependence of double-peak ZBA${\rm {\bf s}}$}\label{Sec:SOMdataBdepNONLIN}

\begin{figure}[h!]
\centering
\includegraphics[width=12cm]{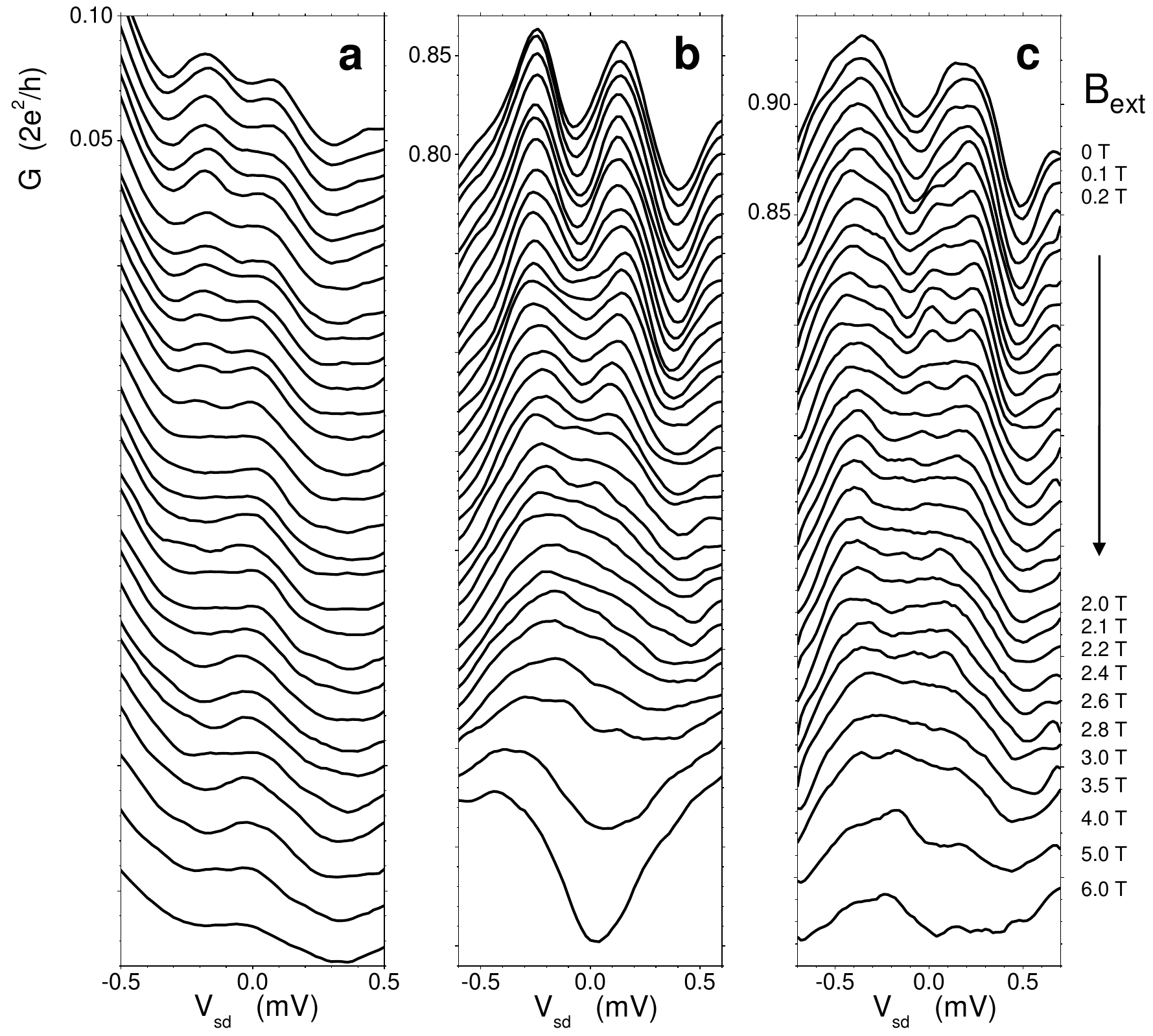}
\caption{Magnetic-field dependence of nonlinear conductance with double-peak ZBAs.
\textbf{a}, Evolution of the nonlinear conductance as a function of applied in-plane magnetic field $B_{ext}$,
for the QPC$_{{\rm 6F}}$ of Fig.~\ref{Fig:FigZBAcharact}c and Fig.~\ref{Fig:FigTBdep}, operated at fixed $V_{g1}=-0.646~{\rm V}$ and $V_{g2}/V_{g1}=0.1$ (giving $L_{eff} \approx 220 ~{\rm nm}$).
Subsequent traces (offset $-0.01 \cdot (2e^2/h)$) are for increasing
$B_{ext}$ from 0 to 2.2~T in steps of 0.1~T, with additional traces for an increase in $B_{ext}$ in larger steps as labeled.
\textbf{b}, As panel \textbf{a}, for the device operated at fixed $V_{g1}=-0.403~{\rm V}$ and $V_{g2}/V_{g1}=0.6$ (giving $L_{eff} \approx 377 ~{\rm nm}$).
\textbf{c}, As panel \textbf{a}, for the device operated at fixed $V_{g1}=-0.398~{\rm V}$ and $V_{g2}/V_{g1}=0.6$ (giving $L_{eff} \approx 377 ~{\rm nm}$).}
\label{Fig:FigBdepNONLIN}
\end{figure}

Three additional examples for the magnetic-field dependence of double-peak ZBAs
are presented in Fig.~\ref{Fig:FigBdepNONLIN}. The dependence on field shows diverse behavior that includes the merging of the two peaks into a single broad peak, after which in some cases a revival of the double-peak character can be observed at higher fields (panels \textbf{a,b}). Panel \textbf{c} presents an example where one of the peaks of the double-peak ZBA at zero field develops a splitting before the full ZBA evolves into a single broad peak.

\newpage

\section{Analysis of ZBA peak positions at ${\bf G=0.4}$, 0.6, 0.7 and ${\bf 0.85 \cdot (2e^2/h)}$}\label{Sec:SOMpeakPosZBA}

\begin{figure}[h!]
\centering
\includegraphics[width=7cm]{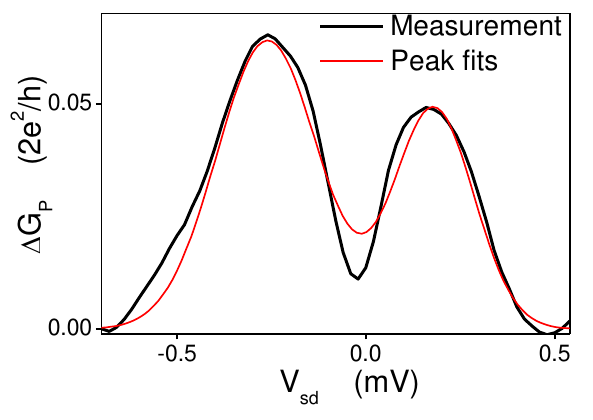}
\caption{Analysis of ZBA peaks. An example of results from fitting two Gaussian peak shapes on a double-peak ZBA. Fits are carried out on peak traces $\Delta G_{P}$, that are obtained by subtracting a background conductance level (linear, or parabolic where needed).} \label{Fig:FigFitting}
\end{figure}

\begin{figure}[h!]
\centering
\includegraphics[width=13cm]{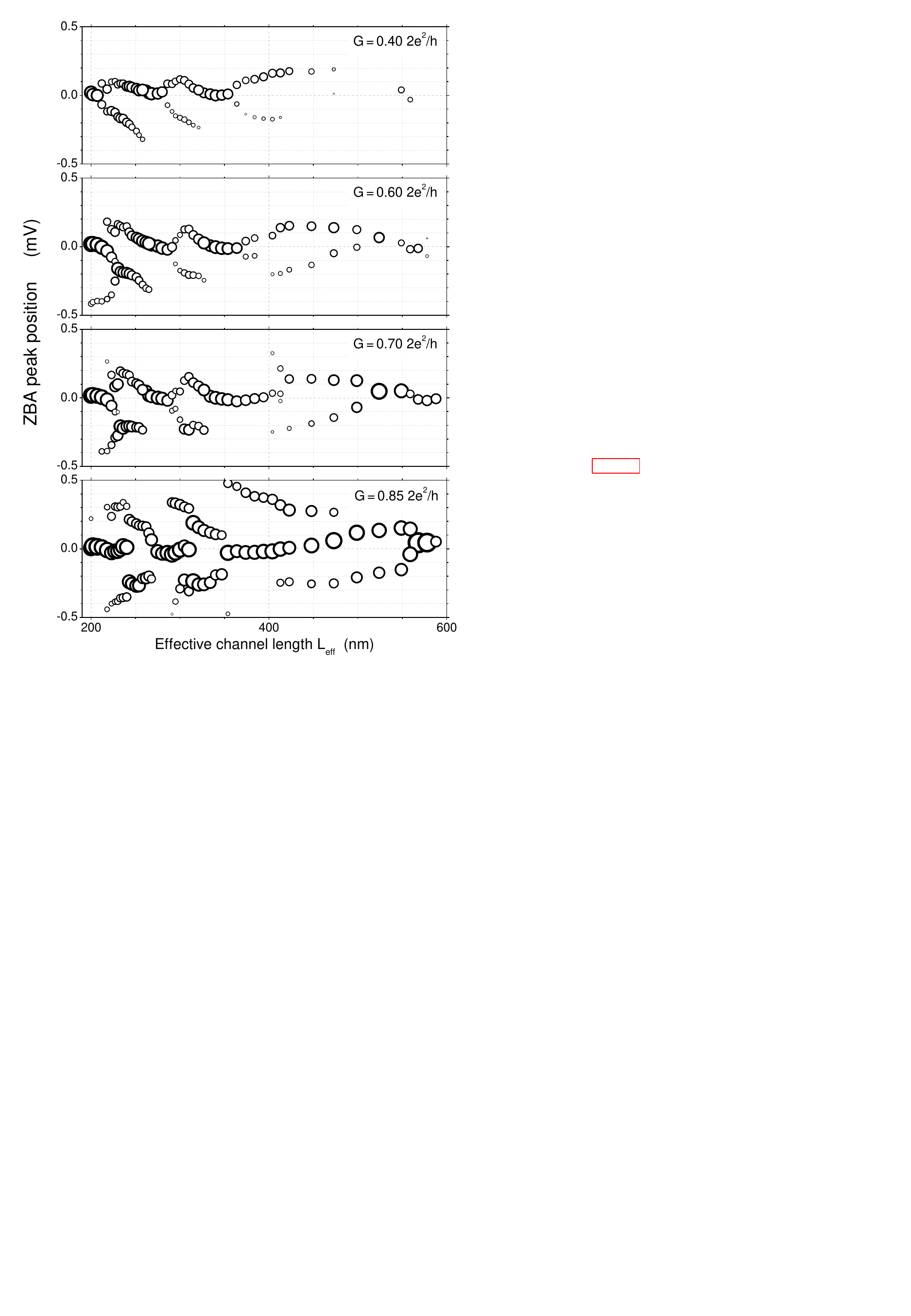}
\caption{Analysis of ZBA peak positions at $G=0.4$, 0.6, 0.7 and $0.85 \cdot (2e^2/h)$ (as labeled). The graphs present
results of fitting ZBA peaks, displayed as peak positions (in $V_{sd}$ units) as a function of $L_{eff}$. The area of the symbols is proportional to the peak area (obtained as the product of peak height and FWHM from peak fitting).}
\label{Fig:FigZBAposAll}
\end{figure}

A detailed analysis of the number of ZBA peaks in nonlinear conductance traces that also quantifies the positions,
widths and amplitudes of these peaks is presented in Fig.~\ref{Fig:FigZBAposAll}.
This analysis was carried out on the data that underlies Fig.~\ref{Fig:FigZBAcharact}c.
Figure~\ref{Fig:FigZBAposAll} presents results for 4 different conductance levels, as labeled.
The symbol size in Fig.~\ref{Fig:FigZBAposAll} is proportional to peak area, which was obtained from fitting Gaussian peak shapes to the ZBA peaks, see Fig.~\ref{Fig:FigFitting} (a phenomenological \textit{ansatz} suited for extracting values for peak position, width and height).
The largest symbols correspond to a peak area of $50~\mu V \cdot (2e^2/h)$.
We mostly observe that the peak width roughly correlates with peak amplitude, such that similar plots with the symbol size proportional to peak width or peak amplitude roughly give the same picture. For the Full-Width at Half-Maximum (FWHM) of ZBA peaks we mainly find values in the range 0.1 to 0.4~mV (or meV for energy scale) for the conductance levels between 0.4 and $0.85\cdot(2e^2/h)$.

\vspace{3cm}

\newpage

\newpage

\section{Linear conductance results for different QPC$_{\bf {\rm 6F}}$ devices}\label{Sec:SOMlinGotherQPC6F}

\begin{figure}[h!]
\centering
\includegraphics[width=9.5cm]{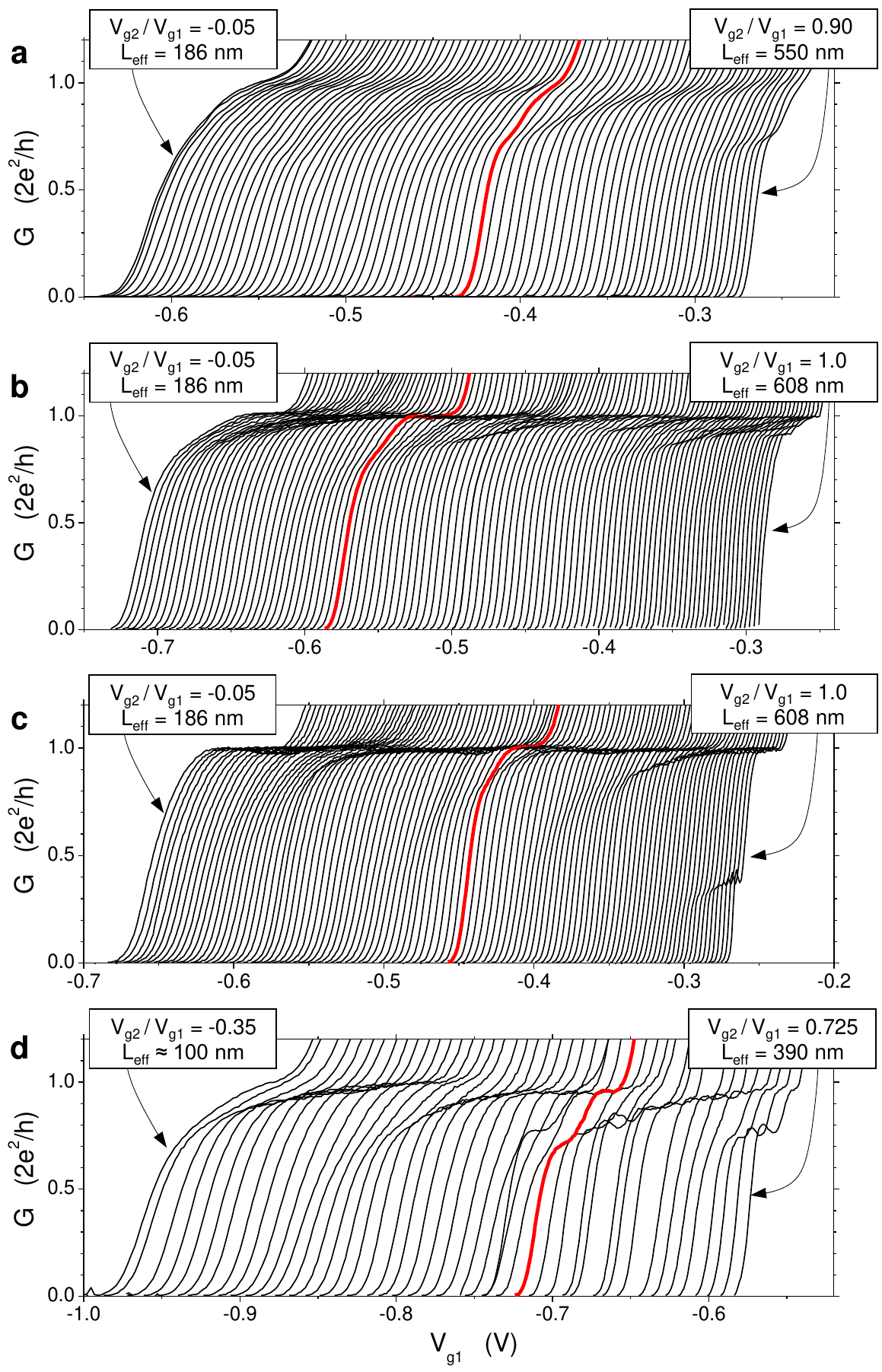}
\caption{Linear conductance $G$ traces as a function of $V_{g1}$ for $L_{eff}$ tuned from a short to a long QPC. The results in \textbf{a-d} are for 4 different devices.}
\label{Fig:Perio4devices}
\end{figure}

Figure~\ref{Fig:Perio4devices} presents linear conductance traces for a range of $L_{eff}$ values from 4 different devices (similar results for again another device were presented in Fig.~\ref{Fig:FigLtuneQPC}c of the main text). These results illustrate that the periodic modulation of the 0.7 anomaly as a function of $L_{eff}$ was observed in all QPC$_{\rm 6F}$ that we measured. The results in Fig.~\ref{Fig:Perio4devices}a-c were obtained on devices where a gate voltage of +0.3~V was applied during cool down. The results in Fig.~\ref{Fig:Perio4devices}d are from a device that had 0~V on the gates during cool down.


\newpage

\section{Full data set for a QPC$_{{\rm {\bf 6F}}}$ device}\label{Sec:SOMfullDataSet}

A full data set of linear and nonlinear conductance for a QPC$_{{\rm 6F}}$ device as a function of $L_{eff}$
is presented in Fig.~\ref{Fig:FullData}, as a sequence of 51 paired graphs.
Here $L_{eff}$ is tuned by adjusting $V_{g2}/V_{g1}$ from 0 to 1 in steps of 0.02.
This data set is the basis for the (zero-field, 80~mK traces in the) results that are presented in Figs.~\ref{Fig:FigLtuneQPC}, \ref{Fig:FigZBAcharact}, \ref{Fig:FigTBdep}, \ref{Fig:FigTdepLIN}, \ref{Fig:FigBdepNONLIN} and \ref{Fig:FigZBAposAll}.


%
%

\vspace{2cm}

\begin{figure}[h!]
\centering
\caption{(figure over next 13 pages) Linear and nonlinear conductance data for a QPC$_{{\rm 6F}}$ device, for $L_{eff}$ tuned from short to long by adjusting $V_{g2}/V_{g1}$ from 0 to 1 in steps of 0.02.}
\label{Fig:FullData}
\end{figure}

\newpage
\begin{figure}[h!]
\centering
\includegraphics[height=22.5cm]{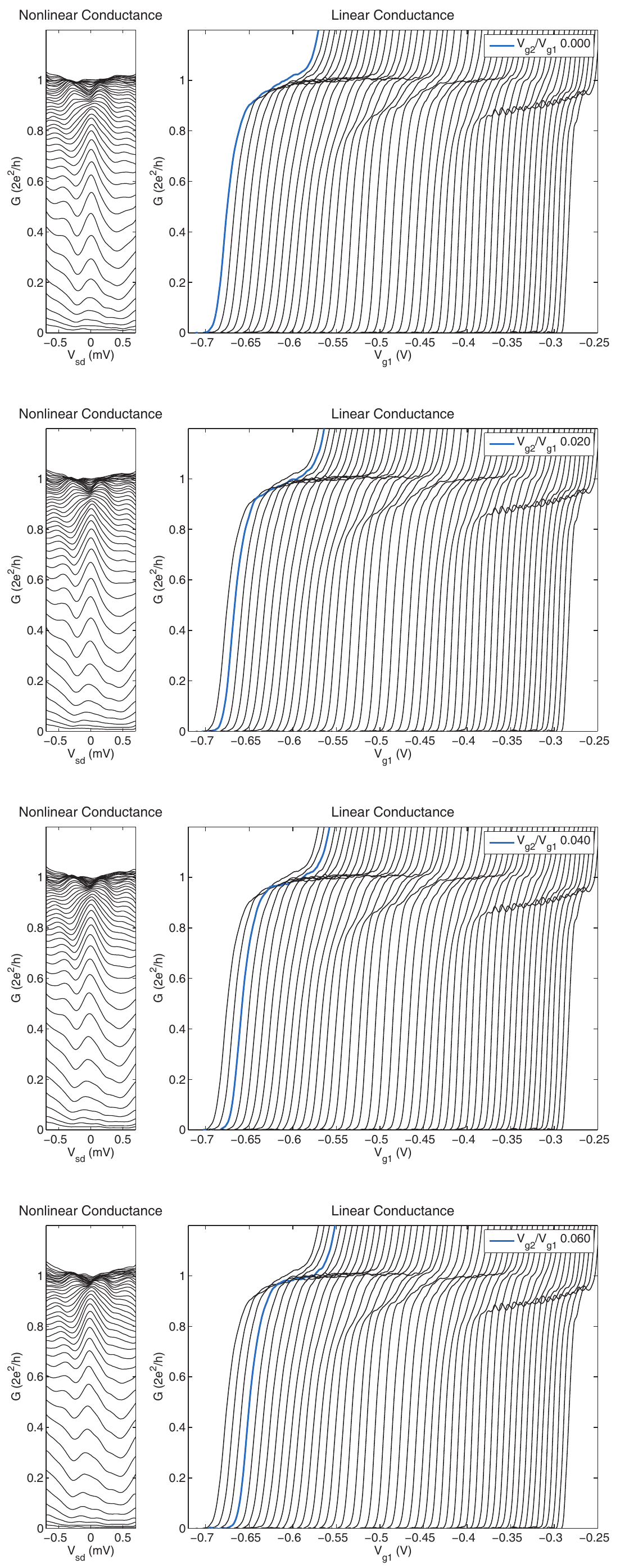}
\end{figure}

\newpage
\begin{figure}[h!]
\centering
\includegraphics[height=22.5cm]{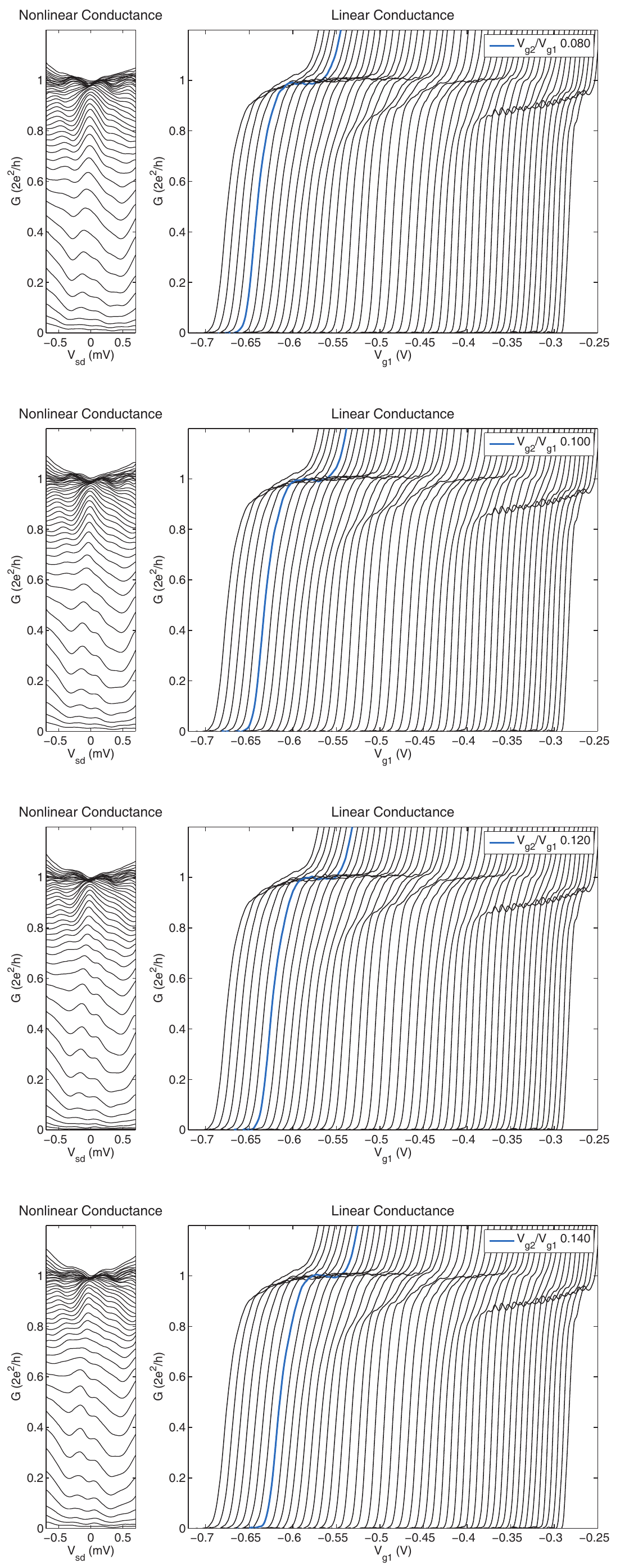}
\end{figure}

\newpage
\begin{figure}[h!]
\centering
\includegraphics[height=22.5cm]{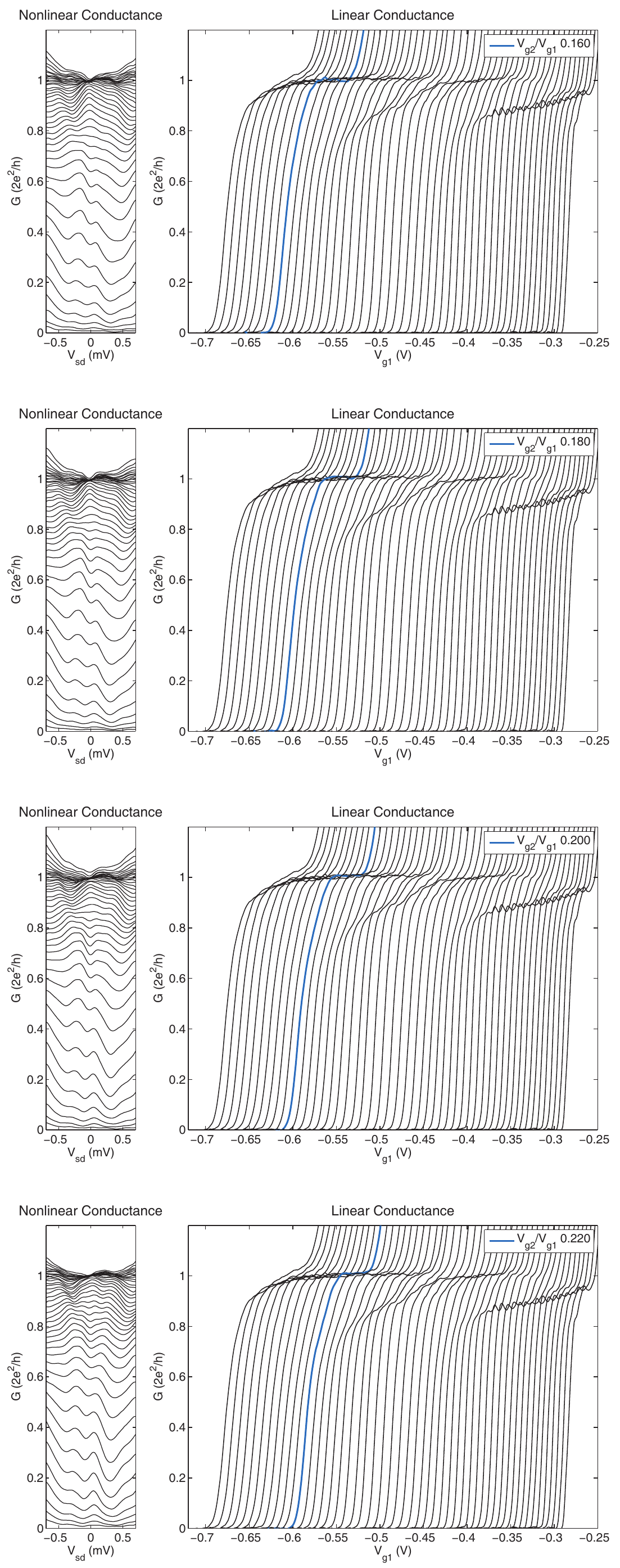}
\end{figure}

\newpage
\begin{figure}[h!]
\centering
\includegraphics[height=22.5cm]{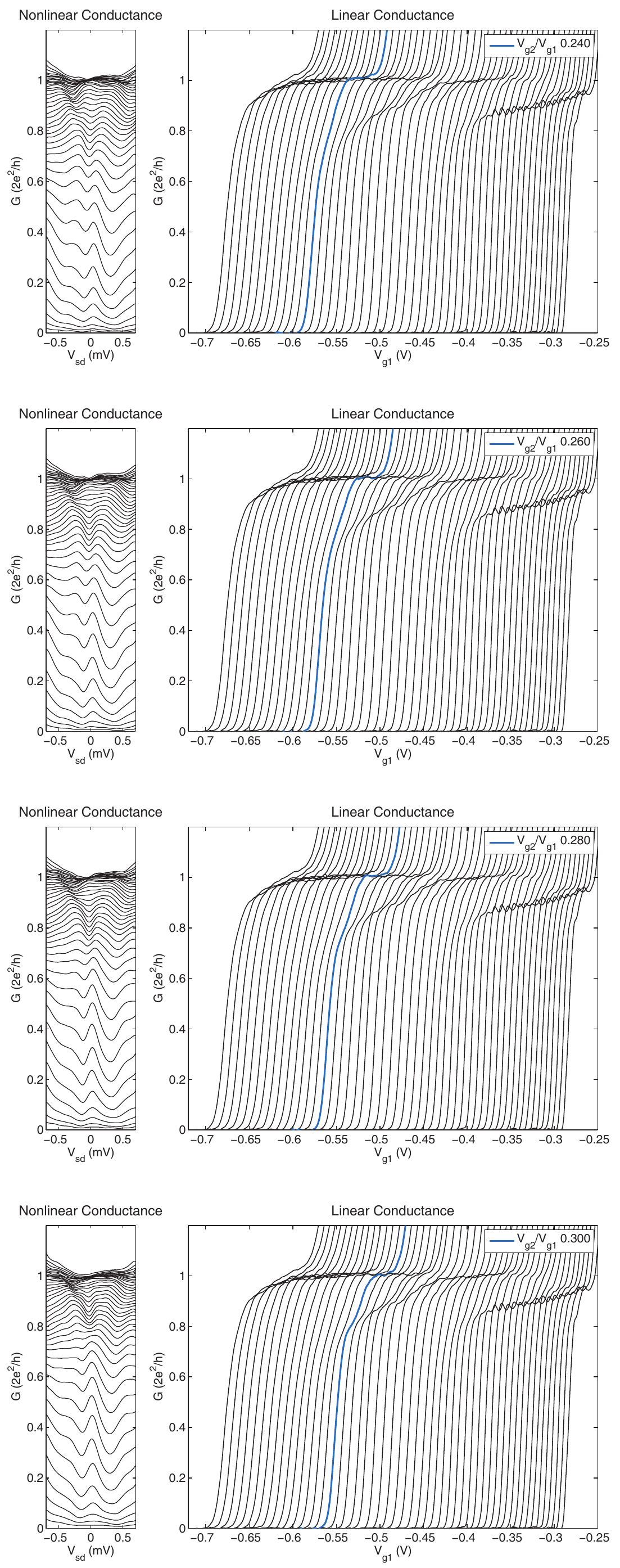}
\end{figure}

\newpage
\begin{figure}[h!]
\centering
\includegraphics[height=22.5cm]{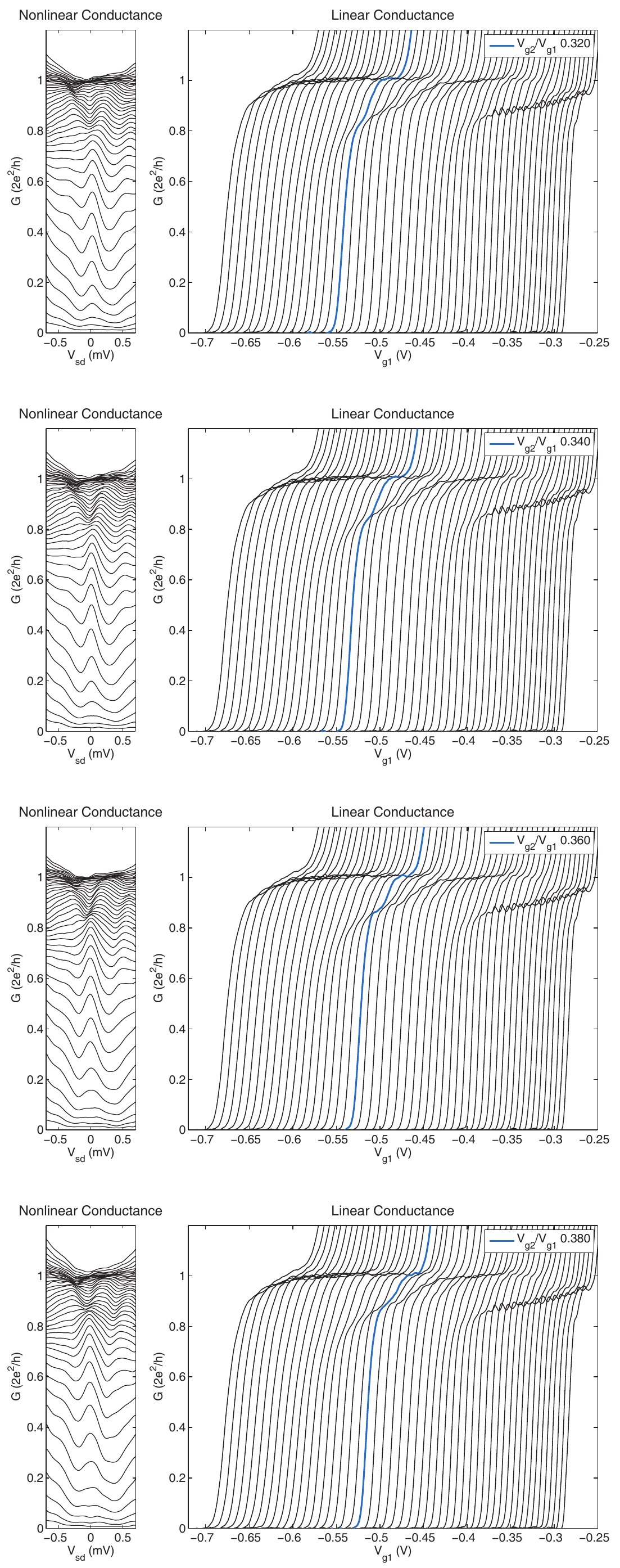}
\end{figure}

\newpage
\begin{figure}[h!]
\centering
\includegraphics[height=22.5cm]{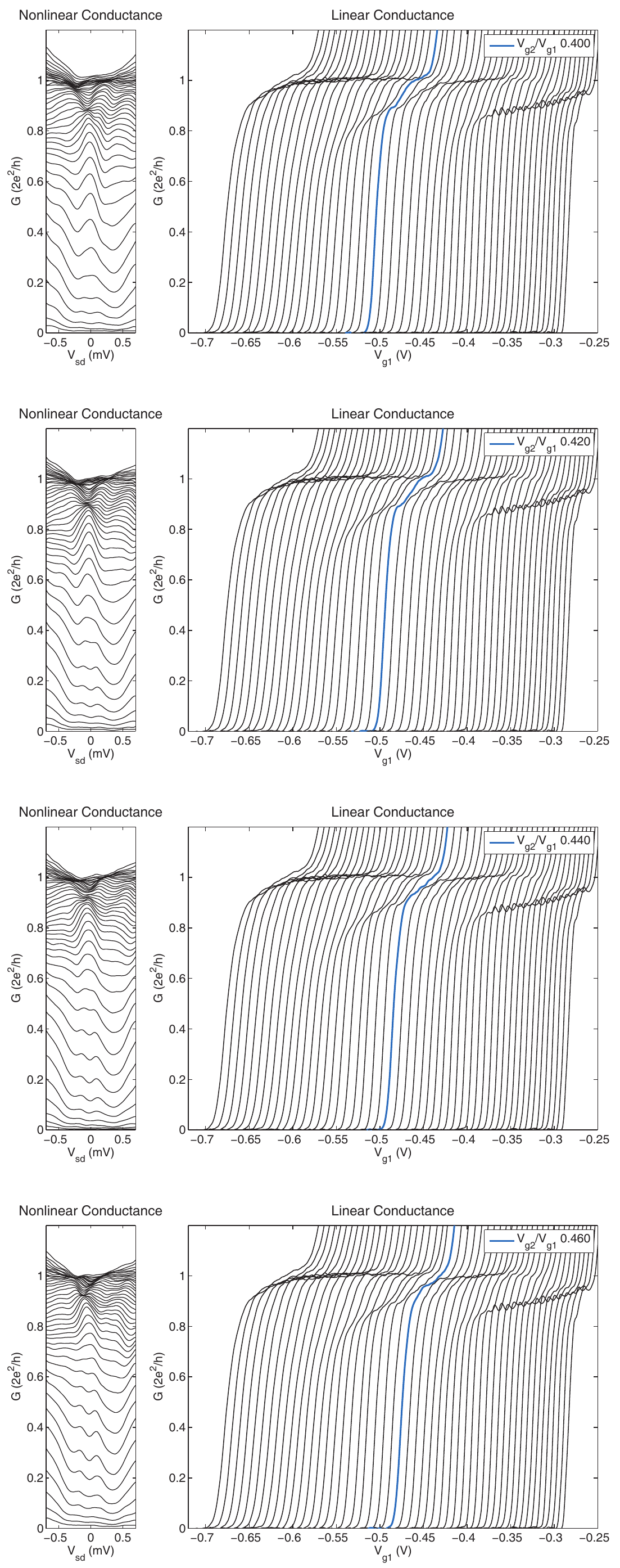}
\end{figure}

\newpage
\begin{figure}[h!]
\centering
\includegraphics[height=22.5cm]{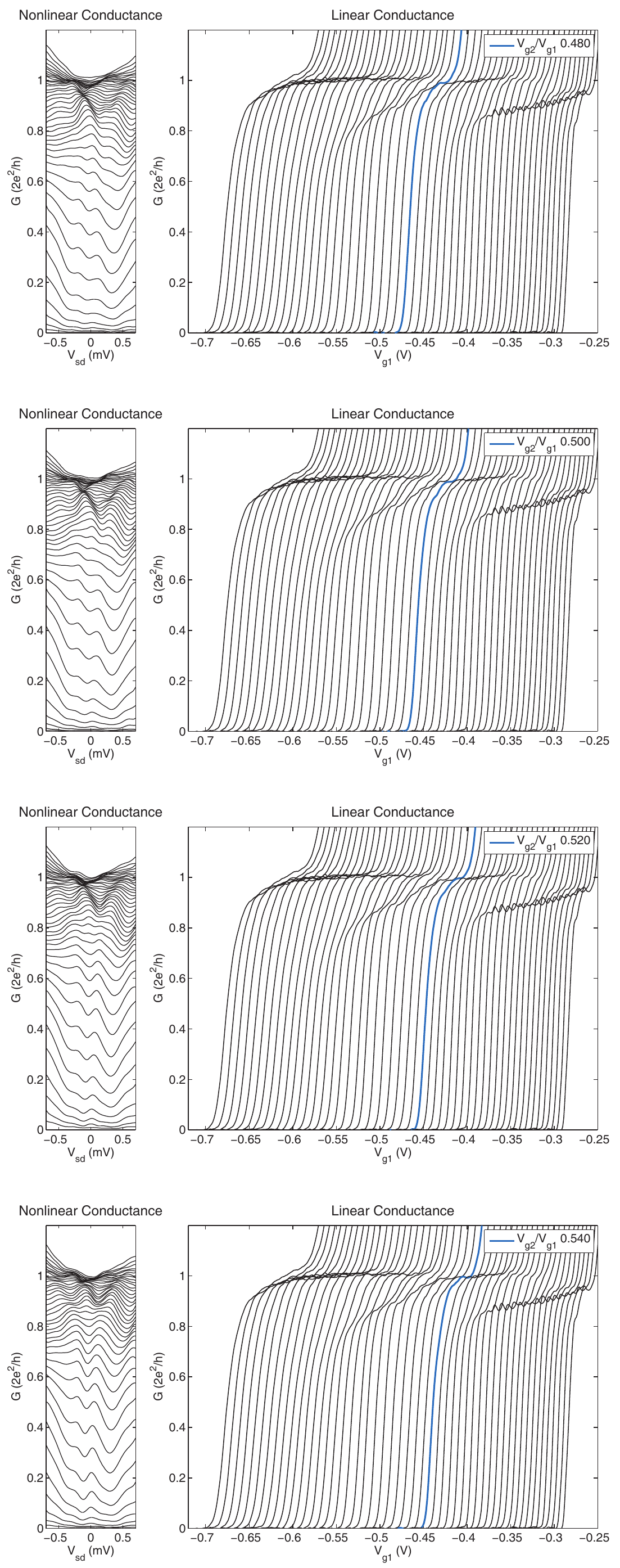}
\end{figure}

\newpage
\begin{figure}[h!]
\centering
\includegraphics[height=22.5cm]{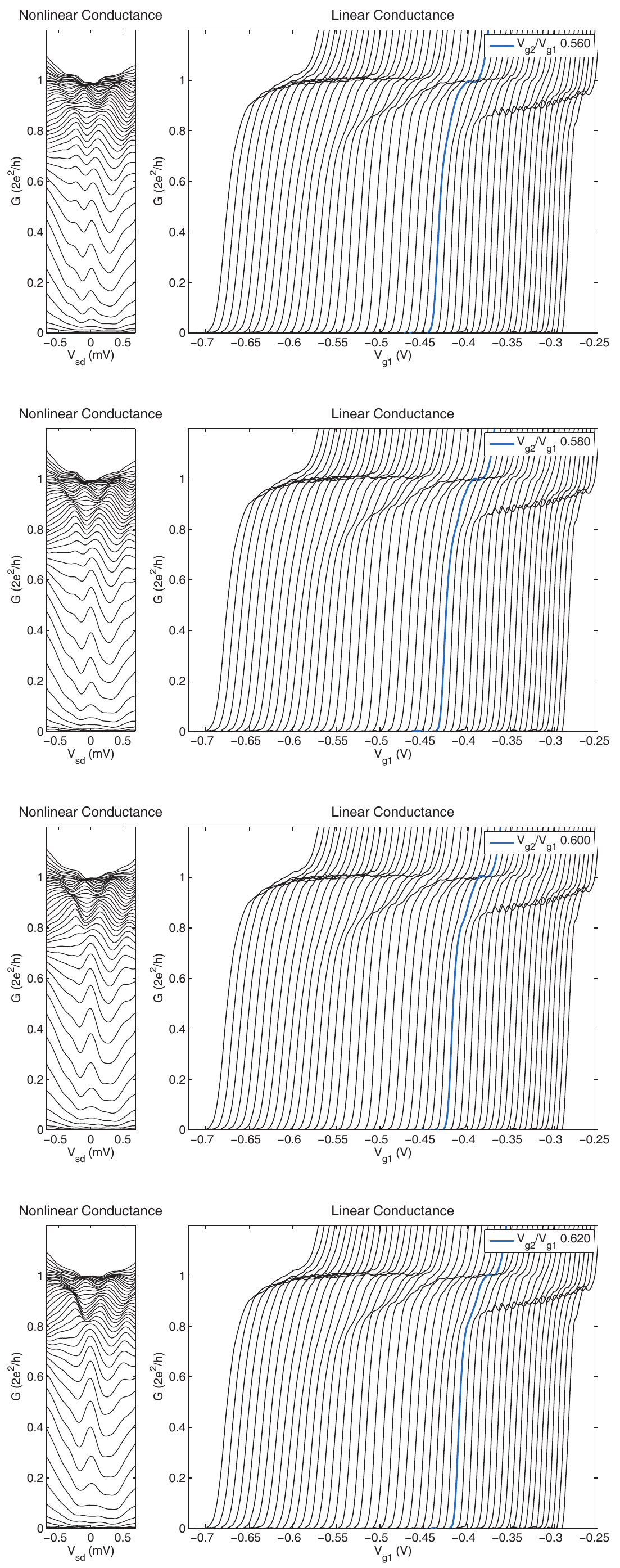}
\end{figure}

\newpage
\begin{figure}[h!]
\centering
\includegraphics[height=22.5cm]{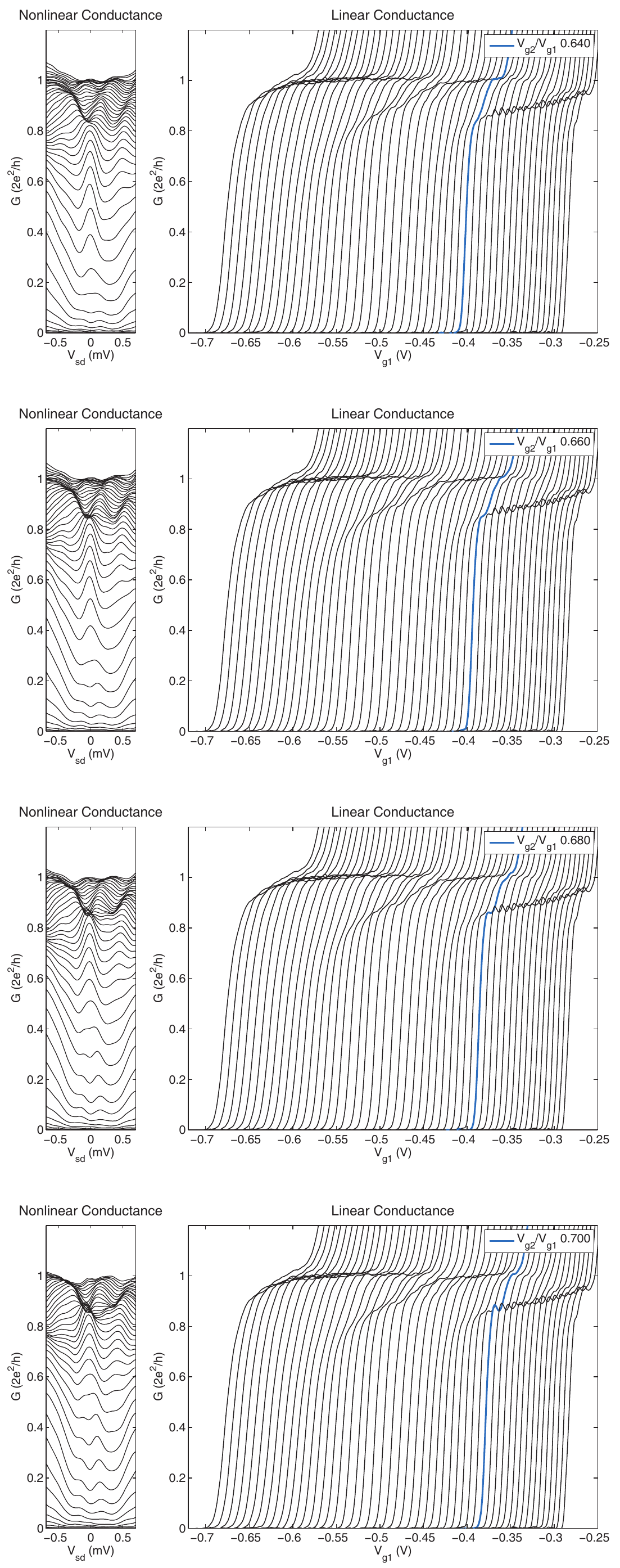}
\end{figure}

\newpage
\begin{figure}[h!]
\centering
\includegraphics[height=22.5cm]{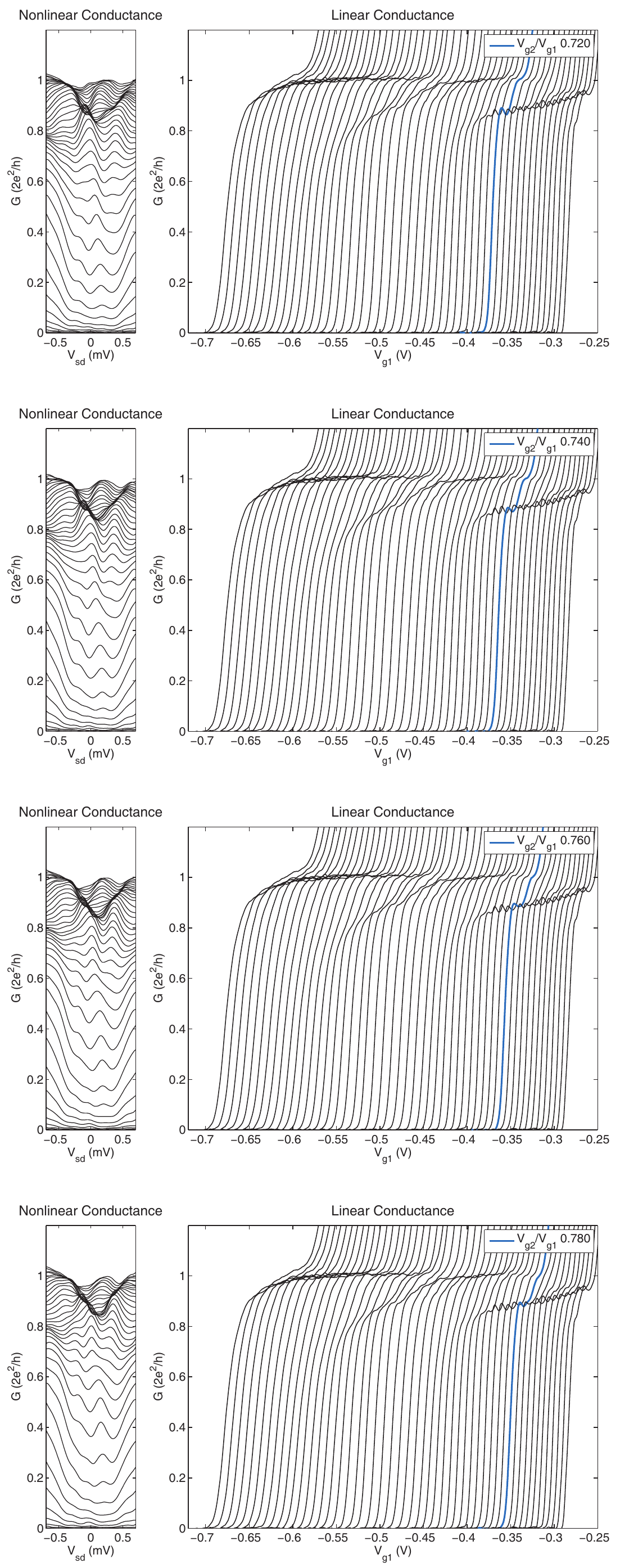}
\end{figure}

\newpage
\begin{figure}[h!]
\centering
\includegraphics[height=22.5cm]{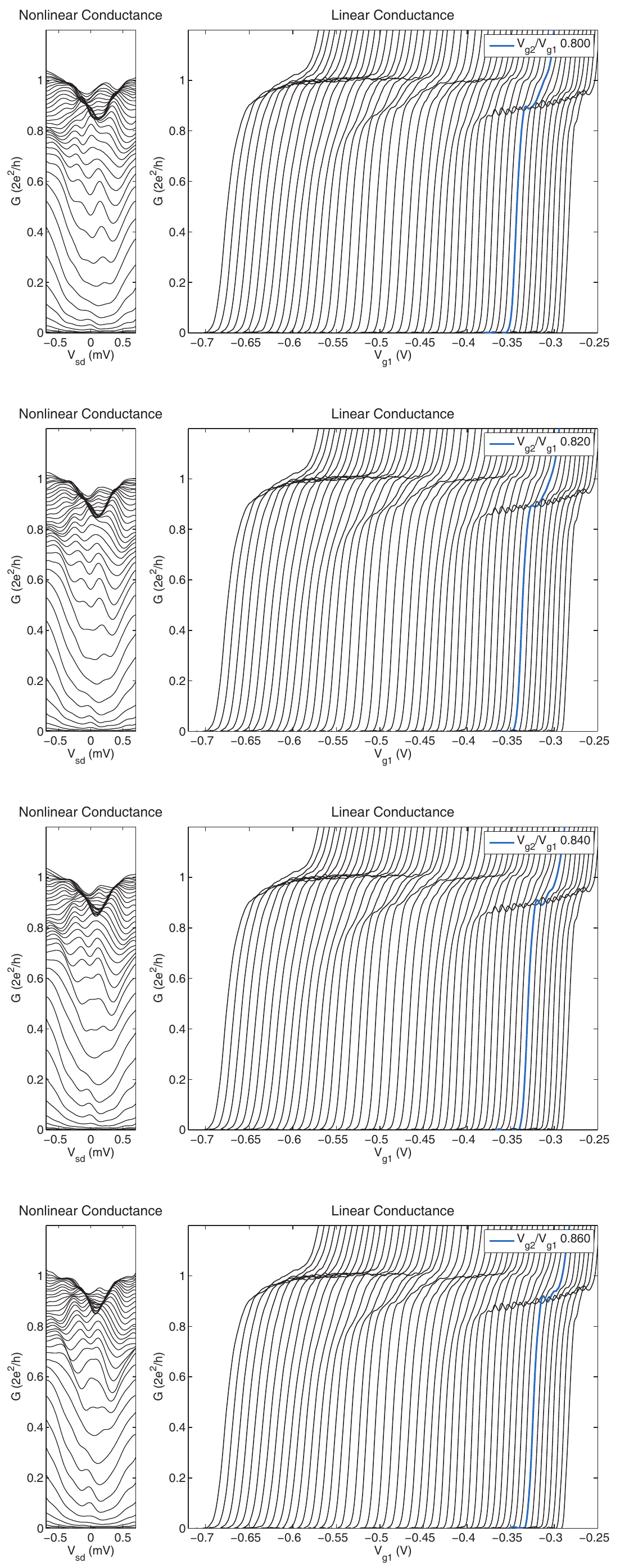}
\end{figure}

\newpage
\begin{figure}[h!]
\centering
\includegraphics[height=22.5cm]{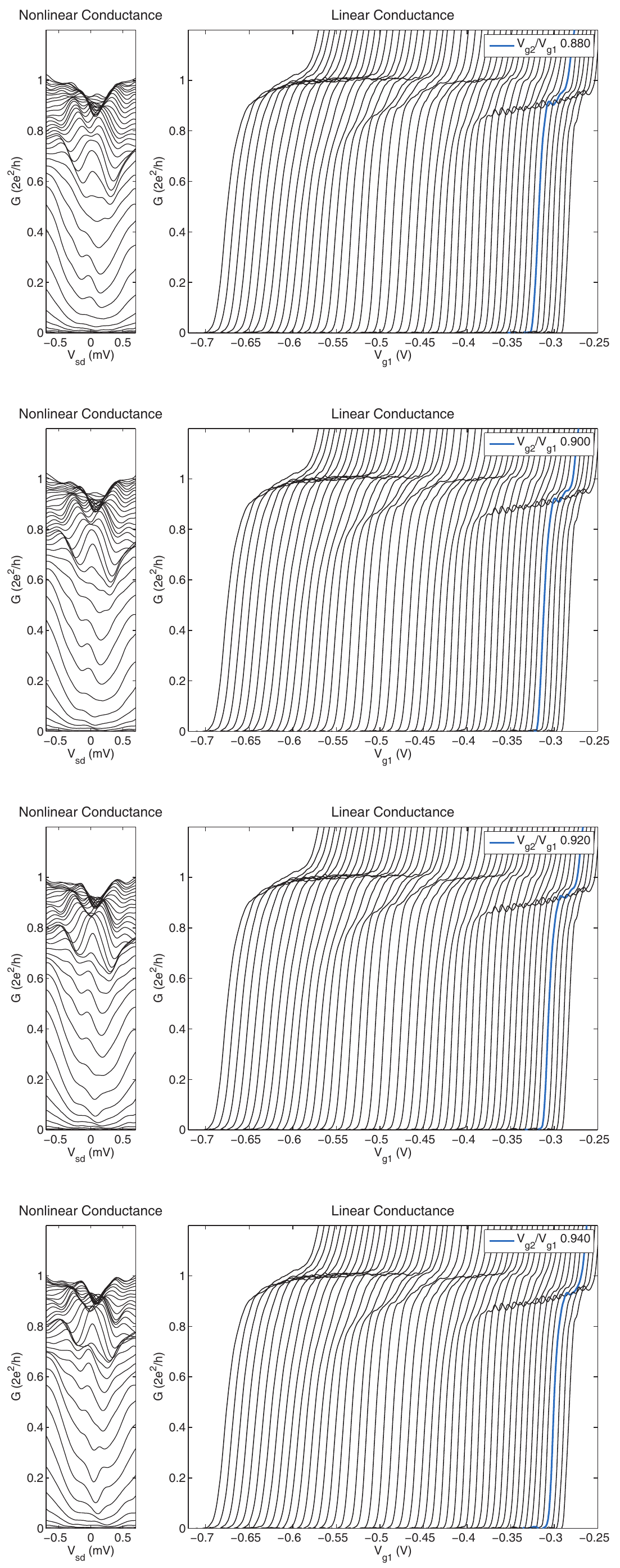}
\end{figure}

\newpage
\begin{figure}[h!]
\centering
\includegraphics[height=22.5cm]{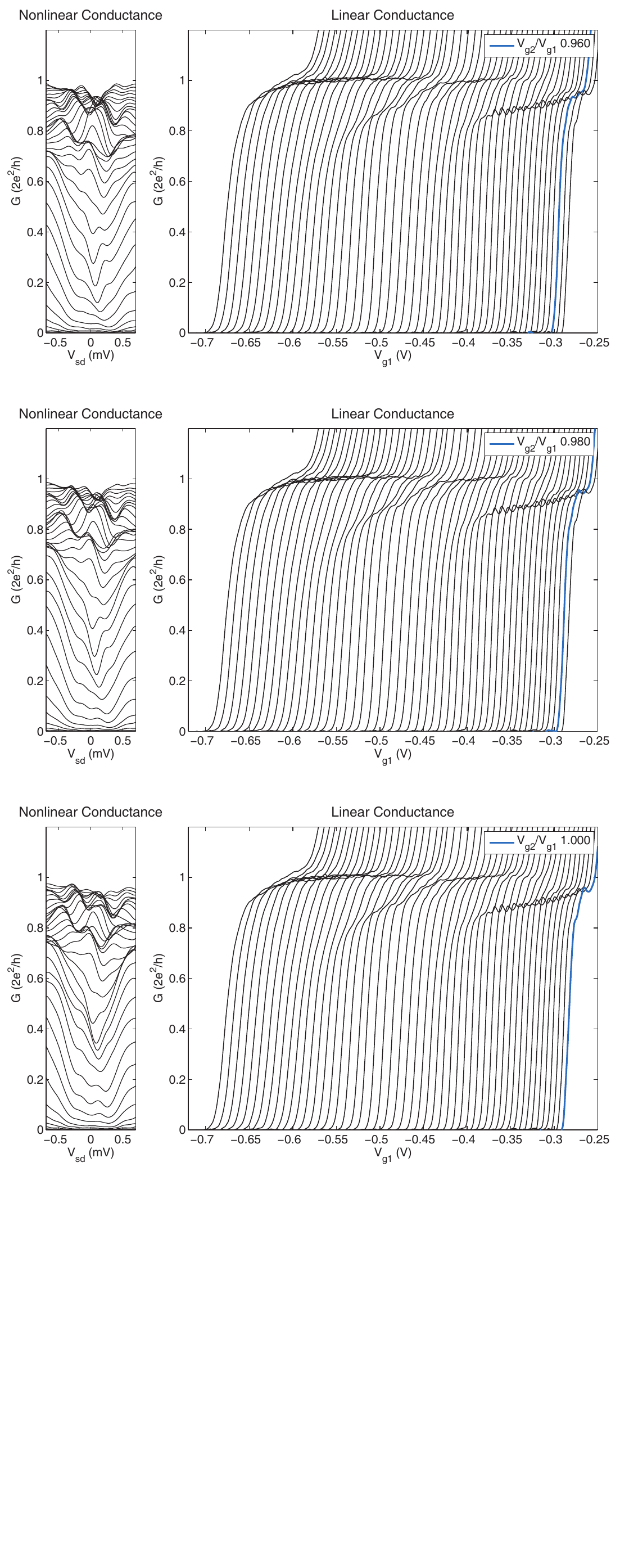}
\end{figure}



\end {document}